\documentclass[a4paper,11pt]{article}
\pdfoutput=1

\usepackage{jheppub}
\bibstyle{JHEP}

\usepackage[english]{babel}
\usepackage[utf8]{inputenc}
\usepackage[T1]{fontenc}
\usepackage{amsfonts}
\usepackage{amsmath}
\usepackage{comment}
\usepackage{tikz}
\usetikzlibrary{decorations.markings}
\usetikzlibrary{decorations.pathmorphing}
\tikzset{snake it/.style={decorate, decoration=snake}}
\usepackage{physics}
\usepackage{dsfont}
\usepackage{nicefrac}

\newcommand{\CP}[1]{\mathds{C}\text{P}^{#1}}
\newcommand{\RP}[1]{\mathds{R}\text{P}^{#1}}
\renewcommand{\i}{\text{i}}
\renewcommand{\d}{\text{d}}
\newcommand{\U}{\text{U}}
\newcommand{\SU}{\text{SU}}
\newcommand{\1}{\mathds{1}}
\newcommand{\tfd}{\text{TFD}}
\newcommand{\p}{{\prime}}

\title{Geometric Phases Characterise Operator Algebras and Missing Information}

\author{Souvik Banerjee, Moritz Dorband, Johanna Erdmenger and Anna-Lena Weigel}

\affiliation{Institute for Theoretical Physics and Astrophysics and Würzburg-Dresden Cluster of Excellence ct.qmat, Julius-Maximilians-Universität Würzburg, Am Hubland, 97074 Würzburg, Germany}

\emailAdd{\{souvik.banerjee, moritz.dorband, erdmenger, anna-lena.weigel\}@physik.uni-wuerzburg.de}

\abstract{We show how geometric phases may be used to fully describe quantum systems, with or without gravity, by providing knowledge about the geometry and topology of its Hilbert space. We find a direct relation between geometric phases and von Neumann algebras. In particular, we show that a vanishing geometric phase implies the existence of a well-defined trace functional on the algebra. We discuss how this is realised within the AdS/CFT correspondence for the eternal black hole. On the other hand, a non-vanishing geometric phase indicates missing information for a local observer, associated to reference frames covering only parts of the quantum system considered.  We illustrate this with several examples, ranging from a single spin in a magnetic field to Virasoro Berry phases and the geometric phase associated to the eternal black hole in AdS spacetime. For the latter, a non-vanishing geometric phase is tied to the presence of a centre in the associated von Neumann algebra.}


\begin{document}
\maketitle
\flushbottom

\section{Introduction}
\begin{figure}[b]
    \centering
    \includegraphics{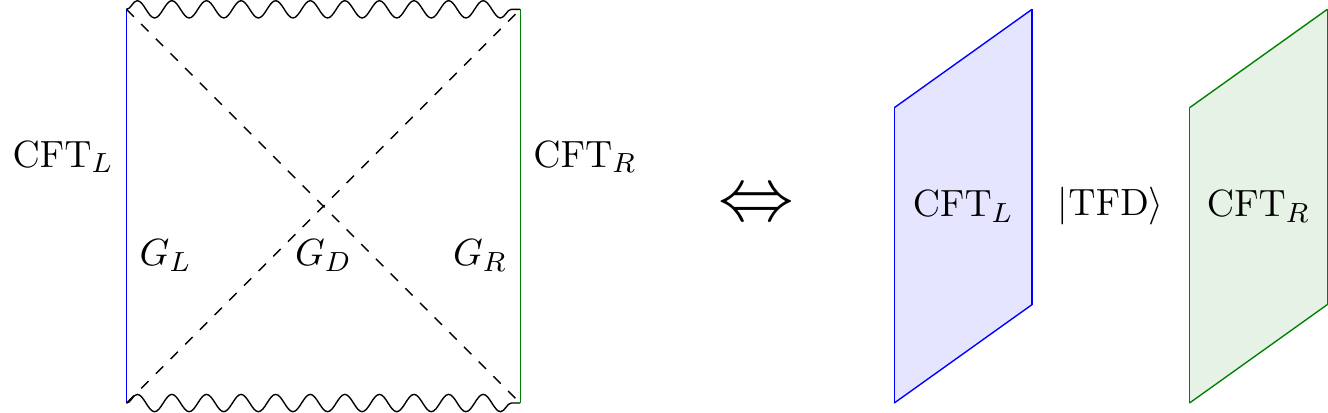}
    \caption{Visualisation of the duality between the eternal black hole in AdS spacetime and the two boundary CFTs entangled in the TFD state \cite{Maldacena:2001kr,VanRaamsdonk:2010pw,Maldacena:2013xja}. On the left-hand side, the eternal black hole in an AdS spacetime is shown in global coordinates. The dashed lines represent the black hole horizons. The wavy lines are the future and past singularities. The two wedges attached to the singularities represent the black hole interior. At the left and right boundaries of the AdS spacetime, marked in blue and green respectively, the left and right boundary CFTs are defined. The left and right asymptotic symmetry groups $G_{L/R}$ constitute global symmetries of the respective CFT. The eternal black hole geometry is invariant under the diagonal subgroup $G_D$ of the full asymptotic symmetry group $G_L\times G_R$. The dual description of the eternal black hole is depicted on the right-hand side. The two CFTs, defined on the blue and green planes that represent the left and right asymptotic boundaries, are entangled in the TFD state.}
    \label{fig:EternalBlackHole}
\end{figure}

The quantisation of gravity is one of the important unanswered questions in theoretical physics. One promising way to approach this problem is provided by the AdS/CFT correspondence \cite{Maldacena:1997re,Witten:1998qj,Gubser:1998bc} as an explicit realisation of the holographic principle \cite{tHooft:1993dmi,Susskind:1994vu}. This correspondence consists of a duality between a theory of gravity in an asymptotically AdS spacetime in $D$ dimensions and a conformal field theory (CFT) without gravity living on its $D-1$ dimensional asymptotic boundary. An example for this duality that will be of particular interest for this paper is the eternal black hole in AdS spacetime, dual to the two CFTs entangled in the thermofield double (TFD) state \cite{Maldacena:2001kr}. These two CFTs are visualised as living on the left and the right asymptotic boundaries of the eternal black hole, as depicted in its Kruskal-Szekeres diagram in fig.~\ref{fig:EternalBlackHole}. The entangled TFD state is built from energy eigenstates of the two CFTs, such that each of the CFTs is thermal with identical temperature fixed by the mass of the black hole. The interior of the eternal black hole is interpreted as a non-traversable wormhole connecting the two boundaries \cite{VanRaamsdonk:2010pw,Maldacena:2013xja}. 

This two-sided geometry is an ideal configuration for analysing the quantisation of a gravity theory, yielding a precise structure of its Hilbert space, the bulk Hilbert space of AdS/CFT. It contains the states corresponding to excitations in both the exterior and the interior of the black hole. A general goal of AdS/CFT, to which also this paper aims at contributing, is to characterise the similarities and the differences between the bulk Hilbert space and the Hilbert space of the dual CFT. This will provide new insights into information processing in a theory of quantum gravity, and a means of contrasting this to the case of a local quantum theory. In this paper, we use the concept of {\it geometric phases} both to reveal the structure of von Neumann (vN) algebras in the given context, and to characterise missing information about the system considered.

\begin{figure}[b]
    \centering
    \includegraphics{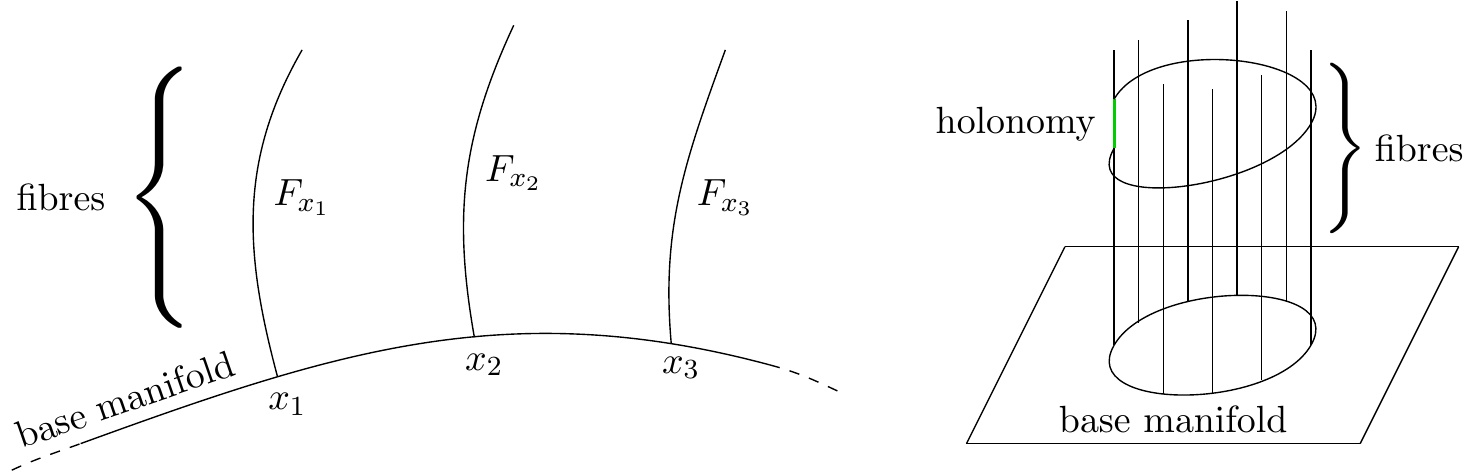}
    \caption{Left panel: a visualisation of the concept of a fibre bundle. At each point $x_i$ of the base manifold, fibres $F_{x_i}$ are attached. For principal fibre bundles, on which we focus in this paper, the fibres are isomorphic to some group $G$. Right panel: a path, closed in the base manifold, may no longer be closed when uplifted in the fibre direction. The mismatch of the endpoints of the uplifted path, marked in green, is the holonomy. In physics, the holonomy is more commonly referred to as geometric phase or Berry phase \cite{Berry:1984jv,Simon:1983mh}. In this work, we use this concept to calculate geometric phases both for small and large spin systems. We connect these results to holonomies of the fibre bundle obtained by quantisation of the moduli space of classical solutions in AdS/CFT.}
    \label{fig:BaseManifoldFibreAndHolonomy}
\end{figure}

To motivate our modus operandi, let us consider the two-sided eternal black hole geometry as in fig.~\ref{fig:EternalBlackHole}. In order to ensure a well-defined variational principle, suitable boundary conditions have to be imposed on the metric. The left and right asymptotic symmetry groups $G_{L/R}$ are defined as the subset of all bulk diffeomorphisms that leave these boundary conditions invariant. The full asymptotic symmetry is then given by $G_L\times G_R$. However, the bulk geometry has an isometry group that corresponds only to the diagonal subgroup $G_D$ of $G_L\times G_R$. The moduli space of classical bulk solutions ${\cal G}_M$ is therefore obtained by quotienting $G_L\times G_R$ with $G_D$. The parameters $g\in{\cal G}_M$ are interpreted as the bulk degrees of freedom. For every choice of parameters $g$, the quantised small fluctuations of $g$ provide Hilbert spaces ${\cal H}_g$. These Hilbert spaces are fibres $F$ in a fibre bundle, with the base manifold $B$ given by ${\cal G}_M$. As visualised on the left-hand side of fig.~\ref{fig:BaseManifoldFibreAndHolonomy}, mathematically, a fibre bundle is defined in terms of four elements $(E,B,\pi,F)$, where $E$ is the total space, $B$ is the base manifold, $\pi$ is a projection from the total space to the base manifold and $F$ are the fibres.

Following the approach of geometric quantisation \cite{Souriau1966geometric,Kostant1970quantization}, the fully quantised Hilbert space is defined as the space of all sections of the bundle. A section corresponds to a choice of local coordinates in the base manifold. Since the quantisation procedure outlined above naturally leads to a fibre bundle structure, an immediate question is whether this bundle is non-trivial. A bundle is trivial when there exists a global section, in which case the full base manifold can be covered by a single coordinate patch. A non-trivial bundle, on the other hand, only allows local sections. Such a non-trivial bundle is quantitatively described by a non-vanishing holonomy that measures whether the endpoints of a closed path initially defined in the base manifold differ by an amount when uplifting the path in the fibre direction. This is visualised by the right panel of fig.~\ref{fig:BaseManifoldFibreAndHolonomy}.
In physics, the holonomy is more commonly referred to as geometric phase or Berry phase \cite{Berry:1984jv,Simon:1983mh}. For the eternal black hole described above, a non-vanishing holonomy is induced by the event horizon. The associated discontinuity of the time-like Killing vector corresponds to a topological defect. From the perspective of the gravitational path integral, this leads to a non-exact symplectic form which is a manifestation of wormhole-like behaviour \cite{Verlinde:2021kgt}. Explicit examples for the physical consequences of non-exact symplectic forms were discussed in \cite{Nogueira:2021ngh,Banerjee:2022jnv}. The topological probes provide a mathematically precise and quantitative description of the thermal behaviour of the black hole, since in this case the holonomy is related to the mass, and correspondingly the temperature of the black hole \cite{Sheikh-Jabbari:2016unm}. At the same time, the black hole temperature is a consequence of the information hidden behind the horizon.

This is an example from gravity for the remarkable more general connection between geometric phases and hidden information.\footnote{A possible relation between geometric phases and hidden information was pointed out to us by Sir Michael Berry during a zoom seminar given by J.E., hosted by GGI Florence in April 2022.} This connection also holds in generic quantum systems without gravity, for example in a system as simple as two coupled spins. This was discussed in \cite{Nogueira:2021ngh} for the case that one of the two coupled interacts with an external magnetic field. It was shown that the factorisation properties of the projective (i.e.~pure-state) Hilbert space are captured by the Berry phase. In a further development \cite{Banerjee:2022jnv}, it was shown that a similar connection holds beyond simple quantum mechanical systems and applies to two-dimensional CFTs as well. In all of these cases, the geometric phases provide a description of wormhole-like structures in the path integral, thus leading to an interesting parallel between information processing in theories with and without gravity. 

In a complementary development, in recent years there has been substantial progress in quantifying the information processing in a black hole spacetime through the understanding of its operator algebra \cite{Leutheusser:2021qhd,Leutheusser:2021frk,Witten:2021unn,Furuya:2023fei}. The categorisation of this von Neumann (vN) algebra determines the process of information scrambling inside a black hole. The main aim of the present work is to find a precise connection {\it between the topological probes discussed above and the defining features of the operator algebra.} In particular, we show how the non-existence of a global section in the fibre structure influences the operator algebra acting on the Hilbert space of a generic quantum system.

We study this influence by carefully analysing the definition of the trace on a vN algebra. The trace is defined as a linear functional $f$, acting on operators $a,b$ of an algebra ${\cal A}$, that is cyclic in its argument,
\begin{align}
    f(ab)=f(ba).\label{eq:Cyclicity}
\end{align}
In quantum theory, a natural definition of an arbitrary linear functional $f^\p$, not necessarily satisfying \eqref{eq:Cyclicity}, is given by the expectation value in a state $|\psi^\p\rangle$. In this paper, we show that an arbitrary linear functional $f^\p$ is proportional to the geometric phase $\Phi_G$ of the state $|\psi^\p\rangle$ when evaluated on the commutator of two operators $a,b\in\mathcal{A}$, $\Phi_G\propto f^\p([a,b])$. Therefore, a state $|\psi\rangle$ with vanishing geometric phase provides a linear functional $f$ that is a well-defined trace, satisfying the required cyclicity property \eqref{eq:Cyclicity}. As we will explain, states with vanishing geometric phase are maximally entangled.

We first obtain this relation using the simple example of two entangled spin $1/2$ particles in a magnetic field. The corresponding operator algebra is of type I.\footnote{For the ease of the reader unfamiliar with vN algebras, we have included a brief summary explaining the aspects relevant for the discussion in this paper in app.~\ref{app:EssentialAspectsvNAlgebras}. For more detailed reviews, we refer the interested reader to \cite{Witten:2018zxz,Witten:2021jzq,Sorce:2023fdx}.} Such an algebra always has an irreducible representation. On the other hand, algebras of type II and type III never have an irreducible representation.  Moreover, a well-defined trace functional only exists for algebras of type I and type II. As mentioned before, the operator algebra for the two-spin system is of type I, so the trace is well-defined. In the approach proposed in this paper, this is reflected by the fact that $f([a,b])$ is proportional to the geometric phase. This phase is given by the symplectic volume of entanglement orbits. These orbits are submanifolds of the projective Hilbert space. Each orbit is associated to a fixed value of entanglement entropy \cite{Sinolecka2002manifolds}. 

The proportionality relation between $f([a,b])$ and the geometric phase becomes even more important when applied to algebras of types II and III. In order to exemplify this, we generalise the aforementioned two-spin model to two copies of infinitely many pairwise entangled spins. We show that if the geometric phases associated with all of these pairs vanish, the associated algebra is of type II, and it is of type III otherwise. While this is a useful connection in any generic quantum theory, we show that there is a further interesting interpretation of these geometric phases for holographic theories. In particular, for time-shifted TFD states in CFT, which are dual to eternal black holes in AdS spacetime, we show that the geometric phase of \cite{Nogueira:2021ngh} is directly related to the centre of the vN algebra. This algebra of type III with non-trivial centre was found to describe the eternal black hole in the large $N$ limit \cite{Leutheusser:2021frk,Leutheusser:2021qhd}. The type II algebra with trivial centre emerges as a consequence of including $1/N$ corrections to the type III algebra \cite{Witten:2021unn}. The connection between the TFD state geometric phase of \cite{Nogueira:2021ngh} and the centre of the algebra of the corresponding holographic CFT exemplifies the proposed {\it differentiation between type II and III algebras in terms of the geometric phase}. The proposed discrimination between type II and type III algebras based on the geometric phase is a novel, universal and fundamental aspect, valid for any quantum theory, with or without gravity.

While the examples mentioned above relate geometric phases to properties of vN algebras quantifying the information processing in an entangled system, geometric phases can be more broadly viewed as quantifying hidden information in any quantum system. This is the second line of thought we develop in this paper giving a fresh perspective on applying  geometric phases to both entangled and non-entangled systems. In both cases, the geometric phases can be thought of as signatures of missing information. This is due to the physical observer accessing only one coordinate patch of the full base manifold. The geometric phase indicates whether the total space $E$ of the fibre bundle is more than simply the product of the base manifold $B$ with the fibre $F$. Provided that the bundle is non-trivial as described above, these phases thus capture if information about the total space is missing when treating the total space as a product in a local description, $E\overset{\text{loc}}{=}B\times F$. We demonstrate this connection using several examples, starting with a single spin in a magnetic field. Geometrically, this example realises the Hopf fibration \cite{hopf1964abbildungen}, where the base space is given by $S^2$ while the fibres are $S^1$. However, the total space $S^3$ is not a simple product of base space and fibre, $S^3\neq S^2\times S^1$, which leads to a non-vanishing geometric phase. As a second example, we also study the Virasoro Berry phase \cite{Oblak:2017ect} based on Virasoro coadjoint orbits \cite{Compere:2015knw} and pinpoint that the missing information is due to non-aligned frames of reference, i.e.~to time-like Killing vectors pointing in different directions. This connection between geometric phase and missing information extends also to entangled systems. We discuss this again using Virasoro Berry phases,  this time in the presence of an eternal black hole. Moreover, we study gauge Berry phases \cite{Nogueira:2021ngh,Banerjee:2022jnv} and modular Berry phases \cite{Czech:2017zfq,Czech:2018kvg,Czech:2019vih,deBoer:2021zlm,Czech:2023zmq}. Although these three kinds of Berry phases are related to different forms of bulk diffeomorphisms,  we find that in all three cases, the origin of the geometric phase is tied to the freedom of different local observers to choose their individual time (or modular time) coordinates.

Our results for the eternal black hole in holography imply an interesting connection between missing information and von Neumann algebras. The non-vanishing geometric phase arises from different coordinate patches for which a local observer in one of them does not have information about the other. At the same time, this geometric phase is tied to the existence of a non-trivial centre for the von Neumann algebra of the eternal black hole. This centre is related to the black hole mass and thus to its temperature.

The paper is organised as follows. We start by briefly reviewing the geometric interpretation of entanglement developed in \cite{Sinolecka2002manifolds} for general bipartite quantum systems. Next in sec.~\ref{sec:TwoSpins}, for illustrative purposes of this construction we explicitly calculate the geometric phase of the coupled two-spin system of \cite{Nogueira:2021ngh}. Subsequently in sec.~\ref{sec:EntanglementTemperature}, we derive the entanglement temperature $T_{\text{ent}}$ for this two-spin system and relate it to the geometric phase. This temperature is defined such that the entanglement entropy mimics the thermal entropy of a system at temperature $T_{\text{ent}}$. With these results at hand, we turn to studying operator algebraic properties of spin systems in sec.~\ref{sec:AlgebraicPerspective}. We first analyse the operator algebra associated with the two-spin system of \cite{Nogueira:2021ngh} for illustrative purposes. In particular, we discuss the definition of the trace and its relation to the geometric phase and the entanglement temperature. Motivated by these results, we generalise our findings to two collections of infinitely many spins with an infinite amount of shared entanglement in sec.~\ref{sec:TraceTypeIITypeIII}. We explicitly show how the values of the geometric phases determine whether a trace can be defined. This also enables us to relate the type of the operator algebra to the values of the geometric phases. Finally in sec.~\ref{sec:RealInHol}, after briefly reviewing the results of \cite{Leutheusser:2021frk,Leutheusser:2021qhd,Witten:2021unn} in sec.~\ref{sec:AlgebraicSetupEternalBlackHole}, we discuss how our result is realised in holography for the eternal black hole in AdS in sec.~\ref{sec:TopPhaseSpaceCentre}. We then move on to study how geometric phases can generically be understood as signalling missing information. We discuss this first for the simple example of a single spin in sec.~\ref{sec:SingleSpin} and second for Virasoro Berry phases in CFT in sec.~\ref{sec:Virasoro Berry phase}. We continue to discuss this for entangled systems in sec.~\ref{sec:EntBerryPhasesInQFT}, starting with Virasoro Berry phases in the presence of a black hole in sec.~\ref{sec:Virasoro Berry phase in the presence of a wormhole}. We then move on to gauge Berry phases in sec.~\ref{sec:GaugeBerryPhase} and modular Berry phases in sec.~\ref{sec:Modular Berry phases}. For the convenience of the reader, we have included a brief review of some aspects of vN algebras in app.~\ref{app:EssentialAspectsvNAlgebras}. In app.~\ref{app:OrbitStructurePureStates}, we provide details on the construction of \cite{Sinolecka2002manifolds}.

\section{Geometric Phase and von Neumann Algebras}
\label{sec:GeomPhaseAndTrace}

In the past years, the concept of geometric phases has played an important role in holography. Different realisations include  modular Berry phases \cite{Czech:2017zfq,Czech:2018kvg,Czech:2019vih,deBoer:2021zlm,Czech:2023zmq}, Virasoro Berry phases \cite{Oblak:2017ect} and geometric phases related to wormhole physics \cite{Nogueira:2021ngh,Banerjee:2022jnv}. In a complementary development, vN algebras have proven useful to obtain new results about holography \cite{Jefferson:2018ksk,Leutheusser:2021frk,Leutheusser:2021qhd,Witten:2021unn} and also for black holes in de Sitter and flat spacetimes \cite{Chandrasekaran:2022cip,Chandrasekaran:2022eqq}.

Motivated by these results, in this paper our goal is to establish a direct connection between geometric phases and vN algebras. As simple examples we study spin systems, in the simplest version consisting of only two spins. In particular, we consider a system of two coupled spins in which one of them also couples to an external magnetic field. For this configuration, the geometric phase of the ground state  was computed in \cite{Nogueira:2021ngh}. We show, first for this two-spin system, that only a state with vanishing geometric phase consistently defines a trace functional on the associated operator algebra. We then generalise this result to operator algebras of type II and type III using two infinite collections of spins. We find that the non-existence of the trace for a type III algebra is due to the fact that every state of the underlying Hilbert space has non-vanishing geometric phase. On the other hand, the Hilbert space underlying a type II algebra does have a state with vanishing geometric phase. We finally discuss how this result is realised in holography for the eternal black hole, combining insights from \cite{Witten:2021unn} and \cite{Nogueira:2021ngh}.

\subsection{Entanglement and Geometric Phase in Bipartite Quantum Systems}
\label{sec:BipartiteQuantumSystems}

We begin our analysis by discussing a relation between geometric phases and entanglement entropy in simple spin systems. A geometric interpretation of entanglement for pure states in $d\times d$-dimensional bipartite quantum systems was established in \cite{Sinolecka2002manifolds}. Since this construction is crucial for the analysis in our paper, we will refer to it as the \textit{SZK construction} throughout this paper. We restrict the present discussion to pure states since defining a trace on an algebra requires using (maximally entangled) state vectors, i.e.~pure states in the language of quantum mechanics.\footnote{As an aside, we point out that a similar analysis for mixed states may be found in \cite{zyczkowski2001monge}.} Our goal in this section is to formulate geometric phases in such systems using the SZK construction
\cite{Sinolecka2002manifolds}. In the following, we only briefly review the construction and state the main results which are immediately important for the discussion of this work.
We include technical details of this construction in app.~\ref{app:OrbitStructurePureStates}. 

The starting point is the Schmidt decomposition of pure states. This decomposition states that for every pure state $|\psi\rangle$ of a bipartite system, a base transformation can be found such that in the new basis, the coefficients $0\leq\kappa_i\leq1$ of the base vectors are the square roots of the eigenvalues of the reduced density matrix,
\begin{align}
    |\psi\rangle=\sum_{i,j=1}^da_{ij}|i,j\rangle=\sum_{i=1}^d\kappa_i|i,\tilde{i}\rangle.\label{eq:SchmidtDecompGeneral}
\end{align}
Alternatively, regarding the $a_{ij}$ as entries of the coefficient matrix $a$, Schmidt decomposition may be phrased as diagonalisation of $a$. The corresponding eigenvalues $\kappa_i$ are known as the Schmidt coefficients of the state.\footnote{Note that, depending on the taste of the authors, sometimes the entries $\tilde{\kappa}_i$ of the reduced density matrix themselves are called Schmidt coefficients, while the state $|\psi\rangle$ is written using $\sqrt{\tilde{\kappa}_i}$, as in \cite{Sinolecka2002manifolds}. This is of course only a matter of convention.} Since they are the square roots of the eigenvalues of the reduced density matrix of each of the $d$-dimensional subsystems, knowledge of all Schmidt coefficients uniquely fixes the value of the entanglement entropy between the two subsystems.

Given a pure state $|\phi\rangle$, there is no measurement that distinguishes between $|\phi\rangle$ and $e^{\i\alpha}|\phi\rangle$. Therefore, these states are physically equivalent. This leads to the definition of the projective Hilbert space ${\cal H}_P$. A state in ${\cal H}_P$ is represented by a ray. A ray consists of all physically equivalent states $\lambda|\phi\rangle$ with $\lambda\in\mathds{C}$ and $\abs{\lambda}=1$. While the canonical Hilbert space of an $N$-dimensional quantum system is $\mathds{C}^N$, the projective Hilbert space of only physically distinct states is given by the complex projective space in one dimension less, $\CP{N-1}$.

The SZK construction \cite{Sinolecka2002manifolds} is based on analysing the Schmidt coefficients and their multiplicities to find the symmetries of a pure state for a fixed value of the entanglement entropy of a $d\times d$-dimensional bipartite quantum system. Based on these symmetries, particular submanifolds of the projective Hilbert space $\CP{d^2-1}$ are associated to particular values of entanglement.\footnote{We point out that the entanglement entropy does not uniquely determine the Schmidt coefficients. Therefore, the submanifolds are, strictly speaking, associated to configurations of Schmidt coefficients rather than values of the entanglement. However, for the special limits of vanishing and maximal entanglement, a particular value for the entanglement uniquely characterises a particular submanifold. Moreover, in secs.~\ref{sec:TwoSpins} and \ref{sec:IllustrativeExampleTwoSpin} we discuss the case of $d=2$ more explicitly, in which due to the simplicity of the system the unique correspondence between the submanifolds and the entanglement holds true for any value of entanglement.} These submanifolds are constructed by quotienting local unitary transformations of the bipartite quantum system, described by $\U(d)\times\U(d)$, with the symmetries of the pure state for a given value of the entanglement entropy. Constructed as homogeneous spaces, the submanifolds can be understood as orbits of the local unitary transformations $\U(d)\times\U(d)$ and are referred to as entanglement orbits.\footnote{As a word of caution, we point out that the phrase `orbit' will also appear later on in sec.~\ref{sec:GeometricPhaseAndMissingInformation} in the context of coadjoint orbits of the Virasoro group. These two concepts of orbits are not to be confused with each other.}

Let us give some examples of such entanglement orbits for a system of two qudits, i.e. two $d$-level systems. In total, there are $d$ Schmidt coefficients that may take values between $0$ and $1$. As discussed above, the projective Hilbert space of each isolated qudit is $\CP{d-1}$. A natural object to construct for a system of two qudits is then the Cartesian product of these two spaces.\footnote{This construction is known as the Segre variety. See e.g. \cite{miyake2003classification,bengtsson2017geometry} for a detailed discussion of this construction in relation to (multipartite) entanglement.} Clearly, this construction will not account for all states of the two-qudit system since the dimension of the space obtained by the Cartesian product is always smaller than the dimension of the projective Hilbert space of the full two-qudit system,
\begin{align}
    \dim(\CP{d^2-1})-\dim(\CP{d-1}\times\CP{d-1})=2(d-1)^2>0\,.
\end{align}
Note that $\dim(\cdot)$ refers to the real dimension. It can be shown that the product space is the orbit which geometrically describes all separable states $|\psi\rangle_{\text{sep}}$ of vanishing entanglement \cite{Sinolecka2002manifolds},
\begin{align}
    |\psi\rangle_{\text{sep}}\in\CP{d-1}\times\CP{d-1}.\label{eq:OrbitVanEnt}
\end{align}
Taking the Cartesian product implies that for separable states, all Schmidt coefficients but one are zero, with the remaining Schmidt coefficient being equal to  $1$. The $d$ different Schmidt coefficients  correspond to the $d$ different possible product states in the two-qudit system, i.e. $|i,\tilde{i}\rangle$ with $1\leq i\leq d$ in the representation of \eqref{eq:SchmidtDecompGeneral}. Geometrically, this reflects that there are $d$ ways to embed $\CP{d-1}\times\CP{d-1}$ into $\CP{d^2-1}$.

Entangled states are obtained by forming linear superpositions of product states. Such states will not be described by the Cartesian product space. In particular, maximally entangled states $|\psi\rangle_{\text{max}}$ belong to an orbit of the form
\begin{align}
    |\psi\rangle_{\text{max}}\in\1\times\frac{\SU(d)}{\mathds{Z}_d}.\label{eq:OrbitMaxEnt}
\end{align}
Here, all Schmidt coefficients are equal and non-zero. This orbit  has an interesting geometric interpretation as well, as first pointed out in \cite{Bengtsson2007Curious}. The dimension of the orbit in this case is exactly half the dimension of the full projective Hilbert space,
\begin{align}
    \dim\Big(\1\times\frac{\SU(d)}{\mathds{Z}_d}\Big)=d^2-1=\frac{1}{2}\dim(\CP{d^2-1})\,.\label{eq:HalfDimension}
\end{align}
Moreover, the symplectic form of $\CP{d^2-1}$ vanishes identically when restricted to the orbit \eqref{eq:OrbitMaxEnt}. This, together with \eqref{eq:HalfDimension}, characterises a Lagrangian submanifold, whose precise definition is as follows. A Lagrangian submanifold $L$ of some manifold $M$ is defined as a submanifold of $M$ with $i)$ the symplectic form $\omega_M$ of $M$ vanishing when restricted to $L$, $\omega_M|_L=0$, and $ii)$ half the dimension of $M$, $\dim(L)=\frac{1}{2}\dim(M)$.\footnote{As an example, a line in $\mathds{R}^2$ is a Lagrangian submanifold. As required, $\dim(\mathds{R})=\frac{1}{2}\dim(\mathds{R}^2)$. The symplectic form of $\mathds{R}^2$, $\omega=\d x\wedge\d y$, vanishes when either $x$ or $y$ are held fixed. As a slightly less simple example, the symplectic manifold of Hamiltonian mechanics spanned by $p_i$ and $q_i$ has Lagrangian submanifolds by fixing either all positions $q_i$ or all momenta $p_i$.}

The intermediate orbits between maximal and vanishing entanglement generically have a more complicated structure. Moreover, with growing value of $d$, there is a large variety of different intermediate geometries, all of which are obtained by considering the corresponding symmetries of states $|\psi\rangle_{\text{intermediate}}$ with intermediate value of entanglement. For instance, if all Schmidt coefficients for a two-qudit state are different but non-zero, the orbit is given by
\begin{align}
    |\psi\rangle_{\text{intermediate}}\in\frac{\U(d)}{\U(1)^d}\times\frac{\SU(d)}{\mathds{Z}_d}.\label{eq:OrbitIntermediate}
\end{align}
The interested reader may consult \cite{Sinolecka2002manifolds} for a detailed list of the intermediate submanifolds up to $d=4$.

The structure of the submanifolds can be interpreted as a fibre bundle \cite{Sinolecka2002manifolds}, with the base space given by the first factors in \eqref{eq:OrbitVanEnt}, \eqref{eq:OrbitMaxEnt} and \eqref{eq:OrbitIntermediate}. The base space consists of all density matrices with the same spectrum. As examples, the base space of the intermediate submanifold \eqref{eq:OrbitIntermediate} consists of density matrices where each eigenvalue appears only once. Such matrices are described by the quotient space $\frac{\U(d)}{\U(1)^d}$. For maximal entanglement, the spectrum of the density matrix is maximally degenerate. The base space of the submanifold \eqref{eq:OrbitMaxEnt} contains only one object, which is the density matrix proportional to the identity.
The fibres of the submanifold arise from pure states. Given a pure state, the corresponding reduced density matrix is obtained by tracing over one of the qudits. Each density matrix has an entanglement spectrum given by the values of the Schmidt coefficients. For a given class of density matrices with the same spectrum, the fibre consists of all pure states leading to a density matrix of that class. For each such submanifold, the connection of the bundle gives rise to a symplectic form. The corresponding symplectic volume has the interpretation of a geometric phase. In the following, we show this for an explicit example, using the approach of \cite{Bengtsson:2001yd} for defining the connection.

\subsubsection{Example for Two Spins}
\label{sec:TwoSpins}

We consider the coupled two-spin model of \cite{Nogueira:2021ngh}. The dynamics of the spins $\vec{S}_i=\frac{1}{2}\vec{\sigma}_i$, where the first spin is under the influence of an external magnetic field, are described by the Hamiltonian
\begin{align}
    H=J\vec{S}_1\cdot\vec{S}_2-2\mu_BBS_{1,z}\,\label{eq:TwoSpinHamiltonian}
\end{align}
with coupling strength $J$ and magnetic interaction strength $\mu_BB$. This system can be interpreted as a hydrogen atom with hyperfine coupling $J$ between the proton and electron spins, $\vec{S}_2$ and $\vec{S}_1$ respectively, in a magnetic field. In first approximation, the Zeeman coupling to the proton spin can be neglected and only the coupling of the electron spin $\vec{S}_1$ to the magnetic field remains. For $J>0$, the ground state of this system is given by
\begin{align}
    |\psi_0\rangle=-\frac{\sin\frac{\alpha}{2}-\cos\frac{\alpha}{2}}{\sqrt{2}}|\!\uparrow\downarrow\rangle-\frac{\sin\frac{\alpha}{2}+\cos\frac{\alpha}{2}}{\sqrt{2}}|\!\downarrow\uparrow\rangle,\label{eq:GroundstateTwoSpin}
\end{align}
where $\tan\alpha=2\mu_B\frac{B}{J}$ and $|i\rangle$ with $i=\,\uparrow,\downarrow$ refer to the first and second spin. From \eqref{eq:GroundstateTwoSpin}, we will obtain the simplest possible realisation of the general geometric SZK construction \cite{Sinolecka2002manifolds} reviewed above, for $d=2$. In this example, the projective Hilbert space is given by $\CP{3}$. To analyse the entanglement properties of \eqref{eq:GroundstateTwoSpin}, we first perform the Schmidt decomposition. As mentioned around \eqref{eq:SchmidtDecompGeneral}, this essentially amounts to a base transformation for the second spin such that \eqref{eq:GroundstateTwoSpin} can be written as $|\psi_0\rangle=\sum_{i=\uparrow,\downarrow}\kappa_i|i,\tilde{i}\rangle$, where $\kappa_i$ are the Schmidt coefficients and $|\tilde{i}\rangle$ is the transformed basis. The new basis is straightforwardly found as $|\tilde{\uparrow}\rangle=|\!\!\downarrow\rangle$ and $|\tilde{\downarrow}\rangle=|\!\!\uparrow\rangle$. The Schmidt coefficients are given by
\begin{align}
    \kappa_\uparrow=\sqrt{\frac{1-\sin\alpha}{2}}\quad\text{and}\quad\kappa_\downarrow=\sqrt{\frac{1+\sin\alpha}{2}}\,.\label{eq:SchmidtCoeffTwoSpin}
\end{align}
With this result, the entanglement entropy between the two spins in the ground state \eqref{eq:GroundstateTwoSpin} is easily computed as
\begin{align}
    S_{\text{EE}}=-\sum_{i=\uparrow,\downarrow}\kappa_i^2\ln\kappa_i^2=\sin\alpha\ln\frac{1-\sin\alpha}{\cos\alpha}-\ln\frac{\cos\alpha}{2}.\label{eq:EntanglementTwoSpin}
\end{align}
Following the  general construction discussed in sec.~\ref{sec:BipartiteQuantumSystems}, depending on the value of $\alpha$, the state \eqref{eq:GroundstateTwoSpin} is part of either $\CP{1}\times\CP{1}$, $\CP{1}\times\RP{3}$ or $\1\times\RP{3}$,\footnote{Note that $\frac{\SU(d)}{\mathds{Z}_d}$ is particularly simple for $d=2$, $\frac{\SU(2)}{\mathds{Z}_2}=\frac{\text{SO}(3)}{\mathds{Z}_2}=\RP{3}$. For general $d$, the orbit of maximal entanglement is not simply a real projective space since $\RP{d^\p}=\frac{\text{SO}(d^\p)}{\mathds{Z}_2}\neq\frac{\SU(d)}{\mathds{Z}_d}$. While relations between $\text{SO}(d^\p)$ and $\SU(d)$ exist, e.g. for $d^\p=6$ and $d=4$, the real projective space requires a quotient by $\mathds{Z}_2$, not by $\mathds{Z}_{d>2}$.} all of which are submanifolds of $\CP{3}$. The explicit construction of these spaces (for a generic two-spin state) is provided in app.~\ref{app:OrbitStructurePureStates}. These spaces are all entanglement orbits that appear in this case.  For $d>2$, there are more intermediate orbits, as pointed out in sec.~\ref{sec:BipartiteQuantumSystems}. In the limit of large $J$, where $\alpha\to0$, the ground state \eqref{eq:GroundstateTwoSpin} is equal to one of the Bell states, i.e.~the maximally entangled states. The entanglement entropy is given by the maximal value  $S_{\text{EE,\,max}}=\ln2$, corresponding to the orbit $\1\times\RP{3}$, cf. \eqref{eq:OrbitMaxEnt}. In the opposite limit of small $J$, where $\alpha\to\frac{\pi}{2}$, the ground state \eqref{eq:GroundstateTwoSpin} becomes a product state $|\psi_0\rangle\sim|\!\!\downarrow\uparrow\rangle$. The entanglement entropy \eqref{eq:EntanglementTwoSpin} vanishes and the corresponding orbit is the Cartesian product $\CP{1}\times\CP{1}$. For all other values of $\alpha$, the orbit is given by $\CP{1}\times\RP{3}$, with the radius of $\CP{1}$ depending on $\alpha$. Note that while all orbits of intermediate entanglement have this geometry, they are still different orbits, distinguished by the value of $\alpha$. In particular, the geometric phase of each such orbit will be different, as we show by explicit computation in the following.

We now compute the geometric phase given by the symplectic volume of the entanglement orbit. The first step is to define a connection on the orbit, following \cite{Bengtsson:2001yd}. To do so, we pick some point $P$ in the orbit, described by the square root of the reduced density matrix. A particularly convenient starting point is to choose the diagonal form of the reduced density matrix with the Schmidt coefficients \eqref{eq:SchmidtCoeffTwoSpin} as diagonal entries. For the two-spin system of the present discussion, this point is given by
\begin{align}
    P=\begin{bmatrix}\sqrt{\frac{1-\sin\alpha}{2}}&0\\0&\sqrt{\frac{1+\sin\alpha}{2}}\end{bmatrix}.
\end{align}
Equivalently, $P$ may be understood as interpreting the Schmidt decomposed version of the ground state \eqref{eq:GroundstateTwoSpin} as an operator acting only on the first spin. This is achieved by the linear transformation $|i,\tilde{i}\rangle\mapsto|i\rangle\langle\tilde{i}|$. Given such a $P$, we then use an arbitrary $\SU(2)$ transformation $u$ to transport $P$ to an arbitrary point $Q$ in the orbit via
\begin{align}
    Q=uP,\quad u=e^{-\i\frac{\phi}{2}\sigma_z}e^{-\i\frac{\theta}{2}\sigma_y}e^{\i\frac{\phi}{2}\sigma_z}.\label{eq:DefQandu}
\end{align}
Since $\tr\big(Q^\dagger Q\big)=\tr\big(P^\dagger P\big)=\tr(\rho_{\text{red}})=1$, differentiating once w.r.t.~the parameters $\phi$ and $\theta$ of the $\SU(2)$ transformation yields
\begin{align}
    \mathfrak{Re}\big[\tr\big(Q^\dagger\d Q\big)\big]=0,\quad\mathfrak{Im}\big[\tr\big(Q^\dagger\d Q\big)\big]\neq0.
\end{align}
Using the standard definition of the connection, the non-vanishing imaginary part can be utilised to define
\begin{align}
    A=\i\tr\big(Q^\dagger\d Q\big)=\frac{\sin\alpha}{2}(1-\cos\theta)\d\phi.
\end{align}
The corresponding field strength of the connection, which is also the symplectic form on the orbit, is then straightforwardly obtained as
\begin{align}
    \Omega=\d A=\frac{\sin\alpha}{2}\sin\theta\d\theta\wedge\d\phi.\label{eq:SympForm}
\end{align}
Performing the integral of $\Omega$ over the full two-sphere, i.e. computing the symplectic volume, we find
\begin{align}
    V_{\text{symp}}=\int\Omega=\frac{\sin\alpha}{2}\,V(S^2)=2\pi\sin\alpha=\Phi_G,\label{eq:SympVolume}
\end{align}
We note that this matches the result for the geometric phase obtained in \cite{Nogueira:2021ngh}, where the expectation value of the gauge connection in the ground state \eqref{eq:GroundstateTwoSpin} was used. Interestingly, the value of $\alpha$, determining the entanglement between the two-spins by virtue of \eqref{eq:EntanglementTwoSpin}, also fixes the symplectic volume of the entanglement orbit. We will comment further on this observation shortly.

\begin{figure}[hb]
    \centering
    \includegraphics[width=0.5\linewidth]{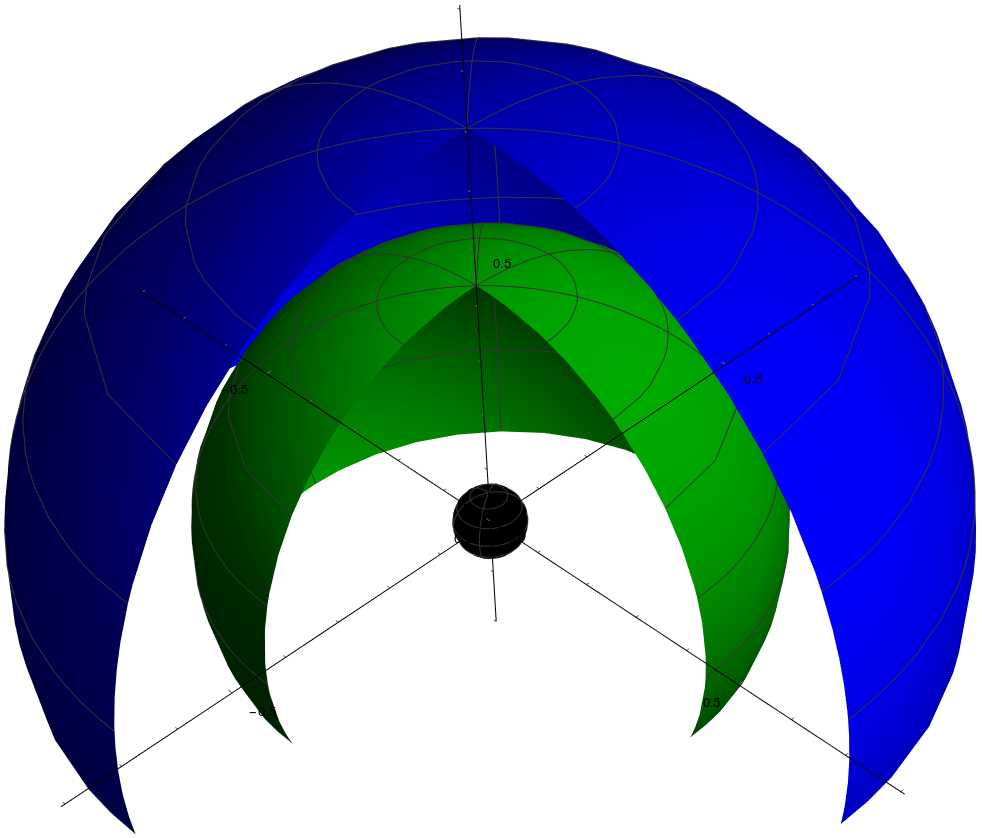}
    \caption{The three differently coloured spheres shown in this plot represent the geometry of the base manifold for three different values of the entanglement. The blue sphere has maximal volume corresponding to $\alpha=\frac{\pi}{2}$ and vanishing entanglement. The green sphere is the base manifold for an intermediate value of entanglement determined by the value of $\alpha$ (for the plot we used $\alpha=\frac{\pi}{8}$). The small black sphere represents the states of (almost) maximal entanglement (in the plot, $\alpha\sim 10^{-3}$).}
    \label{fig:BaseSpaceVisualised}
\end{figure}

The volume \eqref{eq:SympVolume} characterises the number of states within a particular entanglement orbit. For instance, in the case of maximal entanglement, there is only one state resulting in a reduced density matrix proportional to the identity, which is the Bell state. Correspondingly, the geometric phase is minimal (i.e. it vanishes). For intermediate entanglement, the reduced density matrices of two states such as
\begin{align}
    |\psi_+\rangle=\frac{a|\!\uparrow\downarrow\rangle+b|\!\downarrow\uparrow\rangle}{\sqrt{a^2+b^2}}\quad\text{and}\quad|\psi_-\rangle=\frac{b|\!\uparrow\downarrow\rangle+a|\!\downarrow\uparrow\rangle}{\sqrt{a^2+b^2}},\quad\text{where}\quad a\neq b,
\end{align}
have the same eigenvalues and hence the same spectrum. However measurements of the same operator yield different results,
\begin{align}
    \langle\psi_\pm|\sigma_z\otimes\1_2|\psi_\pm\rangle=\pm\frac{a^2-b^2}{a^2+b^2}.
\end{align}
These measurement results become parametrically more and more distinct when moving towards product states, say $a=1$ and $b=0$. Correspondingly, the symplectic volume is larger for small entanglement since there are more states contained in the fibre. The reduced density matrices are general elements of $\SU(2)$ and, for vanishing entanglement, become projectors, $\rho_{\text{prod}}^2=\rho_{\text{prod}}$. Note in particular that, for maximal entanglement, the base manifold only contains the unity element as indicated in \eqref{eq:OrbitMaxEnt}. Geometrically, the factor $\sin\alpha$ in \eqref{eq:SympVolume} is proportional to the radius of the base manifold which, for non-maximal entanglement, is $\CP{1}$. This is visualised in fig.~\ref{fig:BaseSpaceVisualised}.

The observation that larger entanglement corresponds to smaller volume, as shown in fig.~\ref{fig:BaseSpaceVisualised}, may seem somewhat counter-intuitive. A larger volume, i.e. more states in the sense of the previous paragraph, might naively be interpreted as enabling more entanglement between those states. This interpretation does however not coincide with the entanglement we are discussing here. In particular, the volume of the entanglement orbit characterises the number of states with the same amount of entanglement in each state. The volume does not measure any potential entanglement between states in the same orbit. Mathematically, this can be understood as follows. Acting with $u\otimes\1$, $u$ defined in \eqref{eq:DefQandu}, on $|\psi_\pm\rangle$ and computing the reduced density matrix for the first spin results in an expression depending on $\theta$. While the eigenvalues are of course unaltered under this unitary transformation, the $\theta$ dependence reflects that there are many states with the same entanglement. In the case $a=b$ however, where $|\psi_\pm\rangle$ reduce to one of the Bell states, the $\theta$ dependent terms cancel exactly. 

In agreement with this interpretation, we pointed out below \eqref{eq:OrbitMaxEnt} that the orbit of maximal entanglement is a Lagrangian submanifold of the projective Hilbert space. Since the symplectic form of the projective Hilbert space vanishes when restricted to the Lagrangian submanifold, also the symplectic volume vanishes. We will explain in more detail in sec.~\ref{sec:AlgebraicPerspective} how this observation leads to a relation between  a well-defined trace on the algebra and the value of the geometric phase. Within the coupled two-spin system discussed previously, the vanishing symplectic volume arises in the limit of large $J$, for which $\alpha=\arctan(2\mu_B\frac{B}{J})\to0$ in \eqref{eq:SympVolume}. For generic values of $J$ and $\mu_BB$ of \eqref{eq:TwoSpinHamiltonian}, the symplectic volume does not vanish. This can also be understood in terms of entanglement thermodynamics, as we discuss in the next section.

\subsubsection{Entanglement Temperature}
\label{sec:EntanglementTemperature}

Here we establish a relation between the geometric phase $\Phi_G$ given in \eqref{eq:SympVolume} and the effective entanglement temperature $\beta_{\text{ent}}$ induced by the non-vanishing entanglement \eqref{eq:EntanglementTwoSpin}. The entanglement temperature arises from formally interpreting the entanglement entropy as a thermal entropy. For the ground state \eqref{eq:GroundstateTwoSpin}, the reduced density matrix of the first spin\footnote{An analogous argument may be given for the second spin. The only differences are a few signs appearing in intermediate steps of the computation.} is given by
\begin{align}
    \rho_{\text{red}}=P^2=\begin{bmatrix}\frac{1-\sin\alpha}{2}&0\\0&\frac{1+\sin\alpha}{2}\end{bmatrix}.\label{eq:ReducedDensityMatrix}
\end{align}
To study the entanglement thermodynamics, we define the modular Hamiltonian by
\begin{align}
    \rho_{\text{red}}=\frac{e^{-H_{\text{mod}}}}{\tr e^{-H_{\text{mod}}}}.\label{eq:HmodDef}
\end{align}
Since the reduced density matrix \eqref{eq:ReducedDensityMatrix} is diagonal, we make an ansatz for the modular Hamiltonian using the third Pauli matrix,
\begin{align}
    H_{\text{mod}}=h\sigma_z,
\end{align}
where $h$ is an open coefficient to be fixed in the following. Inserting this ansatz into \eqref{eq:HmodDef} and comparing with \eqref{eq:ReducedDensityMatrix}, we obtain an expression for the coefficient $h$ in terms of the Schmidt coefficients,
\begin{align}
    h=\frac{1}{2}\ln\frac{\kappa_\downarrow^2}{\kappa_\uparrow^2}=\frac{1}{2}\ln\frac{1+\sin\alpha}{1-\sin\alpha}.
\end{align}
Inverting the result for the geometric phase \eqref{eq:SympVolume}, we express $h$ by $\Phi_G$,
\begin{align}
    h=\frac{1}{2}\ln\frac{2\pi+\Phi_G}{2\pi-\Phi_G}.
\end{align}
The coefficient $h$ may be interpreted as a magnetic field expressed in units of the temperature. Due to the form of the modular Hamiltonian $H_{\text{mod}}=h\sigma_z$, it can be understood as the Hamiltonian describing a single spin interacting with a magnetic field, with $h$ being the magnetic energy of the spin in the magnetic field, $\mu_BB$, measured in terms of the `thermal' energy $k_BT_{\text{ent}}$.\footnote{Unless stated otherwise, in the following we will work in units where $k_B=1$.} Therefore we find that
\begin{align}
    h=\frac{\mu_BB}{k_BT_{\text{ent}}}=\frac{1}{2}\ln\frac{2\pi+\Phi_G}{2\pi-\Phi_G}\quad\to\quad\beta_{\text{ent}}=\frac{k_B}{2\mu_BB}\ln\frac{2\pi+\Phi_G}{2\pi-\Phi_G}.\label{eq:EntanglementTemperature}
\end{align}
The value of the entanglement temperature is therefore determined by the geometric phase. In particular, in the limit of vanishing entanglement where $\Phi_G\to2\pi$, the logarithm diverges, so the entanglement temperature goes to zero. Furthermore, for maximal entanglement, $\Phi_G=0$. In this case, the logarithm vanishes and accordingly, the entanglement temperature diverges. So we find that by tuning the value of the geometric phase, we cover the full temperature range. Note in particular that the limits $\beta_{\text{ent}}\to0$ and $\beta_{\text{ent}}\to\infty$ exactly reproduce the above behaviour of the entanglement properties when considering the TFD state
\begin{align}
    |\tfd\rangle=\frac{1}{\sqrt{Z}}\sum_ne^{-\beta\frac{E_n}{2}}|n_L\rangle|n_R^\ast\rangle\quad\text{with}\quad Z=\sum_ne^{-\beta E_n},\label{eq:TFD}
\end{align}
which is used to describe thermal systems. For $\beta\to\infty$ (i.e. $T\to0$), the superposition of energy eigenstates within \eqref{eq:TFD} is dominated by the lowest energy, i.e. the ground state. Therefore, the low temperature limit reduces the TFD state to a product state with vanishing entanglement,
\begin{align}
    \lim\limits_{\beta\to\infty}|\tfd\rangle=|0_L\rangle|0_R\rangle\,.\label{eq:TFDLimitLowTemperature}
\end{align}
In the opposite limit $\beta\to0$ (i.e. $T\to\infty$), all of the exponentials in the TFD state $\sim e^{-\beta\frac{E_n}{2}}$ tend to $1$. Therefore, in the high temperature limit the TFD state goes to a maximally entangled state
\begin{align}
    \lim\limits_{\beta\to0}|\tfd\rangle=\frac{1}{\sqrt{\sum_n1}}\sum_n|n_L\rangle|n_R^\ast\rangle.\label{eq:TFDLimitHighTemperature}
\end{align}
This is not a mere coincidence and has an explanation in terms of the geometric interpretation of entanglement analysed above. In calculating the Schmidt decomposition of \eqref{eq:GroundstateTwoSpin}, we have written the state in a way very close to the TFD state,
\begin{align}
    |\psi_0\rangle=\sqrt{\frac{1-\sin\alpha}{2}}|\!\uparrow\!\tilde{\uparrow}\rangle+\sqrt{\frac{1+\sin\alpha}{2}}|\!\downarrow\!\tilde{\downarrow}\rangle.
\end{align}
Expressing this Schmidt decomposed version of \eqref{eq:GroundstateTwoSpin} by the entanglement temperature, we find
\begin{align}
    |\psi_0\rangle=\frac{1}{\sqrt{1+e^{-\beta_{\text{ent}}2\mu_BB}}}\Big[|\!\downarrow\!\tilde{\downarrow}\rangle+e^{-\beta_{\text{ent}}\mu_BB}|\!\uparrow\!\tilde{\uparrow}\rangle\Big],\label{eq:TwoSpinTFD}
\end{align}
which is the TFD state for the two-spin system. This makes manifest the above discussed behaviour of the TFD state in the limits of vanishing and infinite temperature for the two-spin system of \cite{Nogueira:2021ngh}.

\subsection{The Algebraic Perspective}
\label{sec:AlgebraicPerspective}

The above results illustrate how geometric phases arise in bipartite quantum systems. We found that  the geometric phases characterise their entanglement properties. In terms of entanglement thermodynamics, this is explained by the entanglement temperature being determined by the geometric phase.

We note that the SZK construction \cite{Sinolecka2002manifolds}, as already mentioned in sec.~\ref{sec:BipartiteQuantumSystems}, is valid for any dimension $N$ of the bipartite quantum system in question. Thus, we expect that the construction is also valid in the large $N$ limit.  This motivates us to consider also states of type II and type III vN algebras in relation to geometric phases within the SZK construction. Properties that appear irrespective of the particular value of $N$, such as the vanishing geometric phase for maximal entanglement, are expected to be valid in the large $N$ limit as well. The precise structure of the resulting generic entanglement orbits in this limit has not yet been considered. Nevertheless, this will not affect the following discussion.

In particular, we combine the insight that maximally entangled states always have a vanishing geometric phase by the reasoning of \cite{Sinolecka2002manifolds} with the fact that, in the language of vN algebras, maximally entangled states can be used to consistently define a trace on the algebra (see e.g. \cite{Witten:2018zxz}). In the following, we make this connection explicit by showing that only states with vanishing geometric phase can be used to consistently define a trace.

In our analysis, we will not only encounter the more familiar vN algebras of type I, but in particular also the more complicated versions of type II and type III. For ease of the reader unfamiliar with this topic, we have included a brief review of aspects of these algebras important for the following discussion in app.~\ref{app:EssentialAspectsvNAlgebras}. More details can be found in reviews such as \cite{Witten:2018zxz,Witten:2021jzq,Sorce:2023fdx}. For illustrative purposes, in sec.~\ref{sec:IllustrativeExampleTwoSpin} we discuss the two-spin model from above w.r.t.~its algebraic properties. Already at this stage, we will see the relation between the trace on the algebra and the geometric phase. In sec.~\ref{sec:TraceTypeIITypeIII}, we then take the large $N$ limit and discuss collections of infinitely many spins. In this discussion, we will encounter the algebras of type II and type III.

\subsubsection{Illustrative Example: Algebraic Perspective for the Two-spin System}
\label{sec:IllustrativeExampleTwoSpin}

We will now discuss how the value of the geometric phase affects the possibility of consistently defining a trace on the algebra. While our ultimate goal will be to discuss this for algebras of type II and type III, for illustrative purposes it is useful to first make use of the two-spin system again, discussed in sec.~\ref{sec:TwoSpins}. Of course, in this case we are not in a type II or type III scenario: the algebras acting on either of the single spin Hilbert spaces are 2-dimensional and therefore of type I$_2$. The combined algebra, acting on the two spin Hilbert space, is of type I$_4$. The two type I$_2$ algebras commute with each other. Since we are in a type I setting, we know that by definition there must exist a trace on both single spin algebras I$_2$. However, the precise definition will turn out to resemble the cases of type II and type III algebras in close analogy, as we show in the following.

As briefly discussed in the introduction in \eqref{eq:Cyclicity}, a trace on an algebra ${\cal A}$ is defined as a linear functional that is cyclic in its argument. Clearly, an expectation value in some state $|\phi\rangle$ defines a linear functional $f(\cdot)=\langle\phi|\cdot|\phi\rangle$ in that it satisfies
\begin{align}
    f(ca)&=cf(a)\quad\text{for}\quad c\in\mathds{C},a\in{\cal A}\quad\text{and}\\
    f(a+b)&=f(a)+f(b)\quad\text{for}\quad a,b\in{\cal A}.
\end{align}
However, the cyclic property does not hold for generic states $|\phi\rangle$. We may test this explicitly for the two-spin system \eqref{eq:TwoSpinHamiltonian} by defining a linear functional $f_0$ as the expectation value in the ground state \eqref{eq:GroundstateTwoSpin},
\begin{align}
    f_0(a_L)=\langle\psi_0|a_L|\psi_0\rangle.\label{eq:LinFuncTwoSpin}
\end{align}
Here, $a_L$ is any operator of the algebra of the first spin.\footnote{In what follows, the same reasoning also applies for operators of the algebra of the second spin. We only choose operators of the first spin for concreteness and ease of notation.} We borrowed the subscript $L$ in analogy to holographic systems to which we make contact in later sections.

Since the algebra of operators of the first spin is simply I$_2$, it can be represented by the matrix algebra of Hermitian $2\times2$ matrices. The Pauli matrices in combination with the identity matrix provide a convenient basis to parametrise any such matrix. Correspondingly, we may parametrise arbitrary operators as
\begin{align}
    a_L=a_{L,n}\sigma_n,\quad b_L=b_{L,n}\sigma_n,\quad n\in\{0,x,y,z\},\quad a_{L,n},b_{L,n}\in\mathds{R},\label{eq:SingleSpinOperators}
\end{align}
where we denote $\sigma_0=\1_2$. With this parameterisation, we can test whether $f_0$ defined in \eqref{eq:LinFuncTwoSpin} is cyclic in its argument by evaluating it on the commutator of $a_L$ and $b_L$. Explicitly, we find
\begin{align}
    f_0([a_L,b_L])=2\i\sin\alpha\big(a_{L,y}b_{L,x}-a_{L,x}b_{L,y}\big).\label{eq:TwoSpinPreTrace}
\end{align}
If this linear functional is to define a trace, the right hand side has to vanish for any two arbitrary operators $a_L$ and $b_L$. Therefore, vanishing of the bracket is not sufficient. We do however have a different way of making the right hand side vanish by setting $\alpha=0$. Thinking back to the discussion in sec.~\ref{sec:TwoSpins}, this means that the state $|\psi_0\rangle$ is maximally entangled. Explicitly, in the limit $\alpha\to 0$, the ground state of the two-spin system \eqref{eq:GroundstateTwoSpin} reduces to one of the Bell states,
\begin{align}
    \lim\limits_{\alpha\to0}|\psi_0\rangle=\frac{|\!\uparrow\downarrow\rangle-|\!\downarrow\uparrow\rangle}{\sqrt{2}}\,.
\end{align}
So we find that a trace on the algebra I$_2$ can only be consistently defined by a state which has maximal entanglement. This argument can be generalised to I$_d$ and even other types of algebras, as we will discuss shortly. First however, note that due to the normalisation of \eqref{eq:GroundstateTwoSpin}, the trace we have defined by $f_0$ has the somewhat unusual property that the trace of the identity is equal to 1, $f_0(\1_2)=1$, while usually one normalises the trace such that the trace of the identity is equal to the dimension. In fact, for finite dimensional systems, defining a linear functional cyclic in its argument actually only defines an object proportional to the (usual) trace. In the present case of finite dimensional matrix algebras, the proportionality factor is easily fixed by demanding that $\tr(\1_d)=d$. We may achieve this by a simple rescaling of $f_0$,
\begin{align}
    \tr_{\text{I}_2}(\cdot)=2\lim\limits_{\alpha\to0}\langle\psi_0|\cdot|\psi_0\rangle\,.\label{eq:TraceI2}
\end{align}
As mentioned before, the above reasoning can be generalised straightforwardly to higher dimensional systems, such as two qudits. Each of the qudits has a Hilbert space on which an algebra of type I$_d$ acts, with the two copies of I$_d$ commuting. Any state in the combined Hilbert space can be represented in Schmidt decomposition
\begin{align}
    |\psi_d\rangle=\sum_{i=1}^d\lambda_i|i_L,i_R\rangle\,,
\end{align}
and any operator of the `left' qudit algebra as
\begin{align}
    F_d=F_{d,L}\otimes\1_R=\sum_{m,n,k=1}^df_{mn}|m_L,k_R\rangle\langle n_L,k_R|\,,\label{eq:QuditOperator}
\end{align}
with $f_{mn}=f_{nm}^\ast$ to ensure that $F_d$ is Hermitian. Equivalently, in the spirit of above, $F_{d,L}$ may be represented as $F_{d,L}=\tilde{f}_0\1_d+\sum_{i=1}^d\tilde{f}_i\gamma_i$, where $\gamma_i$ span the Lie algebra of $\SU(d)$. Using \eqref{eq:QuditOperator}, a straightforward calculation then shows that
\begin{align}
    f_d(F_d)&=\langle\psi_d|F_d|\psi_d\rangle=\sum_{i=1}^d\lambda_i^2f_{ii}\quad\text{and}\\
    f_d([F_d,G_d])&=\langle\psi_d|[F_d,G_d]|\psi_d\rangle=\sum_{i,j=1}^d\lambda_i^2(f_{ij}g_{ji}-f_{ji}g_{ij}),
\end{align}
where $G_d$ is defined analogously to $F_d$. As long as $\lambda_i\neq\lambda_j$ for any two Schmidt coefficients, $f_d$ does not satisfy the conditions required to consistently define a trace. In particular, $f_d$ is not cyclic in its argument. If however $\lambda_i=\lambda_j=\frac{1}{\sqrt{d}}$ for every $i,j$, i.e. if $|\psi_d\rangle$ is maximally entangled,
\begin{align}
    \langle\psi_d|F_d|\psi_d\rangle&=\frac{1}{d}\sum_{i=1}^df_{ii}\quad\text{and}\\
    \langle\psi_d|[F_d,G_d]|\psi_d\rangle&=0.
\end{align}
As before in \eqref{eq:TraceI2}, by rescaling with $d$, the expectation value in the state $|\psi_d\rangle$ with $\lambda_i=\frac{1}{\sqrt{d}}$ for all $i=1,...,d$ matches the usual notion of the trace in that it satisfies $f_d(\1_d)=d$.

We have discussed above in \eqref{eq:TwoSpinPreTrace} how the entanglement properties of a state in a two-spin system are directly related to the possibility of defining a trace on the algebra, in that the state must be maximally entangled. By our above analysis in sec.~\ref{sec:BipartiteQuantumSystems}, these states have vanishing geometric phase. To make this explicit, we rewrite the result \eqref{eq:TwoSpinPreTrace} for $f_0([a_L,b_L])$ in terms of the geometric phase. Inverting \eqref{eq:SympVolume}, we find
\begin{align}
    f_0([a_L,b_L])\propto\Phi_G\,.\label{eq:TwoSpinTraceGeomPhase}
\end{align}
Since only the maximally entangled states define a trace and those states belong to an orbit with vanishing symplectic volume following our discussion at the end of sec.~\ref{sec:TwoSpins}, the trace can only be defined using states with vanishing geometric phase. This is consistent with our generalisation to qudits with type I$_d$ algebras since, as discussed in sec.~\ref{sec:BipartiteQuantumSystems}, the orbit of states with maximal entanglement is always a Lagrangian submanifold of the projective Hilbert space and thereby always has vanishing symplectic volume.

Whether $f_0$ consistently defines a trace or not can also be understood by the entanglement thermodynamics and in particular the value of the entanglement temperature derived in sec.~\ref{sec:EntanglementTemperature}. In fact, \eqref{eq:TwoSpinPreTrace} can be expressed as
\begin{align}
    f_0(a_Lb_L-b_La_L)\propto\big(1-e^{-\beta_{\text{ent}}\mu_BB}\big),\label{eq:TwoSpinTraceTemperature}
\end{align}
using \eqref{eq:EntanglementTemperature}. In the case of maximal entanglement where $\beta_{\text{ent}}=0$, the r.h.s.~vanishes, consistent with our above result. The r.h.s of \eqref{eq:TwoSpinTraceTemperature} is reminiscent to what happens when discussing the trace for type II and type III algebras by use of the thermofield double state for infinite dimensional spin systems, see e.g. \cite{Witten:2021jzq}. As mentioned before \eqref{eq:TwoSpinTraceTemperature} and discussed in detail in sec.~\ref{sec:EntanglementTemperature}, the entanglement temperature and the corresponding TFD state description \eqref{eq:TwoSpinTFD} of the two-spin ground state \eqref{eq:GroundstateTwoSpin} are in one-to-one correspondence to the geometric phase \eqref{eq:SympVolume} showing up on the r.h.s. of \eqref{eq:TwoSpinTraceGeomPhase}. Motivated by this similarity in structure, we expect a result similar to \eqref{eq:TwoSpinTraceGeomPhase} when discussing algebras of type II and type III explicitly in sec.~\ref{sec:TraceTypeIITypeIII}.

Before we generalise the above result to algebras of type II and type III, we give an at least intuitive reason to why the value of the geometric phase appears when discussing the trace, following the results on the complex quantum geometric tensor of \cite{Provost:1980nc,Alvarez-Jimenez:2017gus}. In quantum mechanical state space, the complex quantum geometric tensor is used to define the metric and the symplectic form on state space \cite{Provost:1980nc}. The metric and the symplectic form are given as the real and imaginary part of the complex quantum geometric tensor, respectively. From the symplectic form, the geometric phase is found by integration. Therefore, the symplectic form is identified with the Berry curvature. In \cite{Alvarez-Jimenez:2017gus}, the notion of this tensor was generalised to field theory by a path integral approach. By considering small variations of parameters $\lambda_a$ that the Lagrangian depends on, the authors define deformation operators $O_a$. The Berry curvature is then (schematically) found to be calculated by an expectation value of the commutator of the deformation operators
\begin{align}
    F_{ab}\propto\langle\Omega|[O_a,O_b]|\Omega\rangle,
\end{align}
where $|\Omega\rangle$ is the ground state.\footnote{The ground state appears because the authors of \cite{Alvarez-Jimenez:2017gus} prepared the path integral such that at past and future infinity, the system is in the ground state. By choosing different boundary conditions, a similar analysis could be performed for excited states, resulting in an analogous formula for the Berry curvature.} By our above results, if $|\Omega\rangle$ is such that it defines a trace, the Berry curvature (and thereby also the geometric phase) has to vanish by the cyclicity property of the trace.

Of course, the approach of \cite{Alvarez-Jimenez:2017gus} is based on field theory which typically has an operator algebra of type III$_1$, as mentioned in the brief review in app.~\ref{app:EssentialAspectsvNAlgebras}. However, as we will show in the following, the relation between the existence of a trace and the value of the geometric phase has a straightforward generalisation to algebras of type II and type III.

\subsubsection{Generalisation to Infinitely Many Spins}
\label{sec:TraceTypeIITypeIII}

We are now ready to generalise our above results to two infinite collections of qubits, as considered in the original construction of the (hyperfinite)\footnote{Hyperfinite refers to the fact that these are constructed as limits of finite dimensional systems. vN algebras appearing in physics can typically be treated as such a limit.} type II and type III vN algebra factors \cite{von1939infinite,powers1967representations,araki1968classification}.

For infinite-dimensional systems, the partition function $Z$ generically diverges in the thermodynamic limit since it is a sum of infinitely many positive numbers. Correspondingly, the thermal density matrix $\rho_{\text{th}}=\frac{e^{-\beta H}}{Z}$ and the underlying Hilbert space are ill-defined. Nevertheless, a proper treatment of physical systems in the thermodynamic limit may still be performed by involving the thermofield double state \cite{haag1967equilibrium}. This state is essentially a purification of the thermal system, achieved by doubling the degrees of freedom and introducing a second copy\footnote{To be precise, the two copies are related by time reversal. However in most discussions of the TFD state, the system at hand is time reversal symmetric. In what follows, our systems will possess the same symmetry.} of the original system. With this state, it is possible to obtain a well-defined Hilbert space by following the GNS construction \cite{gelfand1943imbedding,segal1947irreducible}. In this procedure, the TFD state takes the role of the cyclic and separating vector. This in particular also works for the TFD state as the dual description of the eternal black hole in AdS spacetime \cite{Maldacena:2001kr} used to define the operator algebras acting on the Hilbert space of the eternal black hole \cite{Leutheusser:2021frk,Leutheusser:2021qhd,Witten:2021unn}.

We now analyse the construction of the hyperfinite factors of type II and type III w.r.t.~its geometric phases. For more details on the construction itself, a review may be found in \cite{Witten:2018zxz}. The cyclic and separating vectors used in defining these algebras can be interpreted as the TFD state for two copies of an infinite collection of spins with the temperature given as the entanglement temperature, in the same sense as the TFD state in sec.~\ref{sec:EntanglementTemperature} for the two-spin system. As before, we will refer to these copies as `left' and `right' in the following. Within states, we always treat the left copy as the first entry within $|\cdot\rangle$. 

We start by considering two spins, one each from the left and right copy, to be in a state with non-vanishing entanglement. A convenient way to denote this state is given by
\begin{align}
    |\lambda\rangle=\frac{1}{\sqrt{1+\lambda}}\big(|\!\downarrow\downarrow\rangle+\sqrt{\lambda}|\!\uparrow\uparrow\rangle\big)\quad\text{with}\quad0<\lambda\leq1,\label{eq:EntangledSpinLambda}
\end{align}
where the value of $\lambda$ is in one-to-one correspondence to the amount of entanglement between the spins: $\lambda=0$ corresponds to vanishing entanglement with $|\lambda=0\rangle$ a product state and $\lambda=1$ corresponds to maximal entanglement with $|\lambda=1\rangle$ a Bell state. Since we have two collections of infinitely many spins, in generalising the above state we consider the left and right spins to be pairwise entangled in the same way as \eqref{eq:EntangledSpinLambda}. Each such qubit pair is in a state $|\lambda_n\rangle$, with the index $n$ indicating the state of the $n$th spin pair. To keep the discussion general, we consider each of the spin pairs to share a different non-vanishing amount of entanglement. This is achieved by imposing that $0\neq\lambda_n$. Combining all of these pairwise entangled states, the full state is given by the tensor product of all of them,
\begin{align}
    |\Psi\rangle=\lim\limits_{N\to\infty}\bigotimes_{n=1}^N|\lambda_n\rangle=\lim\limits_{N\to\infty}\bigotimes_{n=1}^N\frac{1}{\sqrt{1+\lambda_n}}\big(|\!\downarrow\downarrow\rangle_n+\sqrt{\lambda_n}|\!\uparrow\uparrow\rangle_n\big).\label{eq:CyclicSeparatingArakiWoods}
\end{align}
This state provides a cyclic and separating vector that may be used to construct, for generic values of $\lambda_n<1$, algebras of type III and corresponding Hilbert spaces on which the algebras act.\footnote{The Hilbert spaces in question are considered to be obtained by the GNS construction. They contain all low energy states obtained by applying operators acting on finitely many of the spins in \eqref{eq:CyclicSeparatingArakiWoods} to this state. The Hilbert space therefore has an at most countably infinite dimensional basis, i.e.~it is separable. The state $|\Psi\rangle$ is contained in this Hilbert space as it is obtained by acting on $|\Psi\rangle$ with the unity operator. Further details can be found e.g.~in \cite{Witten:2021jzq} and references therein.} As first shown in \cite{araki1968classification} and mentioned  in app.~\ref{app:EssentialAspectsvNAlgebras}, it depends on the convergence properties of the sequence of $\lambda_n$ which subclass of type III is constructed. If the sequence converges to some value $0\leq\lambda^\ast<1$, the algebra is of type III$_{\lambda^\ast}$. If it converges to $0$ sufficiently fast, the algebra is actually not of type III, but of type I$_\infty$. The reason is that for sufficiently fast convergence, the partition function is actually finite and the Hilbert space for a single copy of the spin collection can be defined. As explained in more detail e.g.~in \cite{Witten:2021jzq}, such a sufficiently fast convergence is present e.g.~if $\lambda_n\sim n^{-\gamma}$, $\gamma>1$, for large $n$. If the sequence does not converge but rather alternates between (at least) two values $0<\lambda^\ast,\tilde{\lambda}^\ast<1$, the algebra is of type III$_1$.

In view of discussing the existence of a trace, we point out that for the state \eqref{eq:EntangledSpinLambda}, a geometric phase and an entanglement temperature can be defined in the same way as discussed in sec.~\ref{sec:TwoSpins} and sec.~\ref{sec:EntanglementTemperature}, respectively. The state \eqref{eq:EntangledSpinLambda} is already written in the Schmidt decomposition. The Schmidt coefficients can be read off directly and are given by
\begin{align}
    \kappa_{\uparrow,n}=\frac{\sqrt{\lambda_n}}{\sqrt{1+\lambda_n}}\quad\text{and}\quad\kappa_{\downarrow,n}=\frac{1}{\sqrt{1+\lambda_n}}
\end{align}
for every of the spin pairs. As shown in sec.~\ref{sec:EntanglementTemperature}, the entanglement temperature follows from the Schmidt coefficients as
\begin{align}
    \frac{\mu_BB_n}{k_BT_{\text{ent}}}=h_n=\frac{1}{2}\ln\frac{1}{\lambda_n}=\frac{1}{2}\ln\frac{2\pi+\Phi_G^{(n)}}{2\pi-\Phi_G^{(n)}}\quad\to\quad\beta_{\text{ent}}=\frac{k_B}{2\mu_BB_n}\ln\frac{2\pi+\Phi_G^{(n)}}{2\pi-\Phi_G^{(n)}}.\label{eq:EntTempLambda}
\end{align}
We choose to let the magnetic energy $\mu_BB_n$ vary for every spin pair, while $T_{\text{ent}}$ does not depend on $n$. While this choice does not affect the discussion of the type of the corresponding algebra, it does yield a more intuitive physical picture. All of the spin pairs feel the same `temperature' while being under the influence of different values of the magnetic field, in close analogy to the TFD state for infinitely extended spin chains.

To give a concrete physical example, we may think of infinitely many copies of the system defined by the Hamiltonian \eqref{eq:TwoSpinHamiltonian} analysed in detail in sec.~\ref{sec:TwoSpins}. The above discussion follows through in exactly the same way by replacing $|\lambda_n\rangle$ with the ground state \eqref{eq:GroundstateTwoSpin}. Allowing for each of the spins of \eqref{eq:TwoSpinHamiltonian} to be under the influence of a parametrically different magnetic field $B_n$ as in \eqref{eq:EntTempLambda}, the parameters map as
\begin{align}
    \lambda_n=\frac{1-\sin\alpha_n}{1+\sin\alpha_n}\quad\text{with}\quad\tan\alpha_n=\frac{2\mu_BB_n}{J}.
\end{align}
Following the same steps for computing the symplectic volume as in sec.~\ref{sec:TwoSpins}, the corresponding geometric phases for the spin pairs are given by
\begin{align}
    \Phi_G^{(n)}=2\pi\frac{1-\lambda_n}{1+\lambda_n}.\label{eq:GeomPhaseLambda}
\end{align}
Hence we may write \eqref{eq:EntangledSpinLambda} for every spin pair using the corresponding geometric phase only, 
\begin{align}
    |\lambda_n\rangle=\frac{1}{\sqrt{2}}\Bigg[\sqrt{1-\frac{\Phi_G^{(n)}}{2\pi}}|\!\uparrow\uparrow\rangle_n+\sqrt{1+\frac{\Phi_G^{(n)}}{2\pi}}|\!\downarrow\downarrow\rangle_n\Bigg].\label{eq:EntangledSpinGeomPhase}
\end{align}
This will be more convenient below.

Next we analyse the existence of a trace on the algebra. As for the two-spin system in sec.~\ref{sec:IllustrativeExampleTwoSpin}, we first need to define a linear functional. In analogy to the case of two spins, we do so utilising the state \eqref{eq:CyclicSeparatingArakiWoods},
\begin{align}
    {\cal F}(a_L)=\langle\Psi|a_L|\Psi\rangle.\label{eq:LinFuncInfSpin}
\end{align}
Here $a_L$ is, as before, any operator acting on the left system of spins. For the infinite collection of spins in the present discussion, such operators may be thought of as finite polynomials of Pauli matrices. For now, we assume that $\lambda_n<1$. For generic values of $\lambda_n$ however, the functional defined above will never be cyclic in its argument for arbitrary operators $a_L,b_L$. To make this explicit, we proceed analogously as in sec.~\ref{sec:IllustrativeExampleTwoSpin}. We parametrise operators acting on a single spin pair by Pauli matrices as in \eqref{eq:SingleSpinOperators}. Then we take a tensor product of finitely many (for concreteness, $k$) of such operators to obtain operators acting on $k$ spin pairs. Evaluating the commutator of two such operators in the linear functional \eqref{eq:LinFuncInfSpin} results in
\begin{align}
    {\cal F}\big([a_L,b_L]\big)=\sum_{n=1}^k\Phi_G^{(n)}g_1^{(n)}(a_L,b_L)+\sum_{n,m=1}^k\Phi_G^{(n)}\Phi_G^{(m)}g_2^{(n,m)}(a_L,b_L)+\mathcal{O}\Big[\big(\Phi_G^{(n)}\big)^3\Big],\label{eq:CheckCyclicityTypeIII}
\end{align}
where $g_i^{(n)}$ are functions only of the operators $a_L,b_L$ that do not vanish in general, analogously to the bracket on the r.h.s.~of \eqref{eq:TwoSpinPreTrace}. Fixing $k$ to some value, these functions may be calculated straightforwardly. Of much higher importance than their explicit form however is the observation that the linear functional evaluated on the commutator receives contributions from every of the individual geometric phases $\Phi_G^{(n)}$. As indicated, various powers of the geometric phases contribute. If $a_L$ and $b_L$ consist of $k$ single spin pair operators, the highest power will be $\big(\Phi_G^{(n)}\big)^k$. We highlight in particular that, while different powers of $\Phi_G^{(n)}$ contribute, there is no term independent of all of the $\Phi_G^{(n)}$. Therefore, ${\cal F}$ evaluated on the commutator vanishes if and only if all of the geometric phases vanish. As we discussed above, for generic $0<\lambda_n<1$, which is necessary to discuss algebras of type III, the geometric phases are non-zero, cf.~\eqref{eq:GeomPhaseLambda}. We have therefore found a geometric reason for the non-existence of the trace on type III algebras.

So far we did not consider the case $\lambda_n=1$ for one or more of the $\lambda_n$. In this case, at least for infinitely many $\lambda_n=1$, the situation changes significantly. Since in this instance, \eqref{eq:CyclicSeparatingArakiWoods} contains only finitely many states $|\lambda_n\rangle$ with $\lambda_n\neq1$, we may use the action of the algebra to transform $|\Psi\rangle$ into a state where $\lambda_n=1$ holds for all $n$. The resulting state describes two collections of infinitely many spins for which each pair of spins is maximally entangled. Moreover, considering \eqref{eq:GeomPhaseLambda}, every of the spin pairs has vanishing geometric phase. By our result \eqref{eq:CheckCyclicityTypeIII}, this shows that the trace can be consistently defined on the algebra, since every of the individual contributions vanishes. We are thus no longer in a setting which leads to defining an algebra of type III. On the other hand, we are still in a setting of two collections of infinitely many spins with an infinite amount of shared entanglement between the two spin collections. Therefore, the algebra acting on the Hilbert space defined from \eqref{eq:CyclicSeparatingArakiWoods} with $\lambda_n=1$ for all $n$ is of type II (cf. the brief review in app.~\ref{app:EssentialAspectsvNAlgebras}). The subclass of the type II algebra is found by observing that the trace is defined for every possible operator of the algebra. In particular, due to the normalisation of the state \eqref{eq:CyclicSeparatingArakiWoods}, the trace of the identity operator is defined and equals $1$. Therefore the algebra is of type II$_1$.

Apart from II$_1$, there are also algebras of type II$_\infty$. As for type I$_\infty$, the trace for II$_\infty$ is not defined for every operator. In fact, such algebras can be constructed as tensor products of II$_1$ and I$_\infty$. This implies that the trace on II$_\infty$ is given by the product of the traces on II$_1$ and I$_\infty$. The above reasoning for the existence of the trace on II$_1$ using geometric phases can therefore also be used to explain the existence of the trace for II$_\infty$.

\subsection{Realisation in Holography: the Eternal Black Hole}
\label{sec:RealInHol}

In the past sections we have shown that states with non-vanishing geometric phases cannot be used to define a trace on the operator algebra. This in particular enabled us to relate the non-existence of the trace for type III algebras to the fact that no state of the Hilbert space of a type III algebra has trivial geometric phase since all these states are non-maximally entangled. In the following we discuss how this is realised in the holographic setting of the eternal black hole. We start by briefly reviewing the results for the corresponding operator algebras of \cite{Leutheusser:2021frk,Leutheusser:2021qhd} and \cite{Witten:2021unn}. We then explain how the results of \cite{Nogueira:2021ngh,Banerjee:2022jnv} fit into this algebraic description, in particular w.r.t.~the relation between the geometric phase and the type of the operator algebra.

\subsubsection{Algebraic Setup for the Eternal Black Hole}
\label{sec:AlgebraicSetupEternalBlackHole}

Within the AdS/CFT correspondence, the eternal black hole is dual to two CFTs entangled in the TFD state \eqref{eq:TFD}, each of the CFTs located on the left and right asymptotic boundaries of the spacetime \cite{Maldacena:2001kr}. Considering the single-trace operators of one of the CFTs (say, the left one), it was shown in \cite{Leutheusser:2021frk,Leutheusser:2021qhd} that in the large $N$ limit, the single-trace operators describe a generalised free field theory. The corresponding operator algebra ${\cal A}_{L,0}$ is of type III$_1$. The commutant of this algebra ${\cal A}_{R,0}$ is again of type III$_1$ and can be understood as the operator algebra of single-trace operators of the right CFT. These two algebras are factors: they only contain operators with non-trivial commutation relations, so the centre is trivial.

In this construction, the CFT Hamiltonian cannot be included in ${\cal A}_{L,0}$. Both the expectation value as well as the two-point function of the CFT Hamiltonian scale with $N^2$ and diverge in the large $N$ limit. While the divergence of the expectation value may be removed by defining $H_L^\p=H_L-\langle H_L\rangle$, the divergence of the two-point function is still present for $H_L^\p$. Therefore this operator cannot be part of the algebra. To define an operator which exists in the large $N$ limit, dividing $H_L^\p$ by $N$ is sufficient, $U=H_L^\p/N$, since this precisely cancels the $N^2$ scaling of the two-point function. However, the operator $U$ is central to ${\cal A}_{L,0}$ for $N\to\infty$,
\begin{align}
    [U,\mathcal{O}]=\frac{1}{N}[H_L^\p,\mathcal{O}]=-\frac{\i}{N}\partial_t\mathcal{O}\overset{N\to\infty}{\to}0\label{eq:CentralOperator}
\end{align}
for any operator $\mathcal{O}\in{\cal A}_{L,0}$. $U$ may be included into the algebra by a tensor product ${\cal A}_L={\cal A}_{L,0}\otimes{\cal A}_U$, where ${\cal A}_U$ is the algebra of bounded functions of $U$. The analogous argument can be run for $H_R^\p$, with operators chosen from ${\cal A}_{R,0}$. Interestingly, to construct ${\cal A}_R$ the same operator $U$ has to be used. Therefore, the centre of both the left and right algebras ${\cal A}_L$ and ${\cal A}_R$ is shared. Physically, this can be understood by the fact that the mass of the black hole has to be the same when measured from both sides. In this sense, from the bulk perspective, the mass corresponds to a shared degree of freedom. The operator $U$ is defined by a rescaling of the Hamiltonian, the latter one measuring the energy (i.e. the mass).

The shared degree of freedom can also be understood by the fact that the difference of the Hamiltonians $h=H_L-H_R$ is a well-defined operator in the large $N$ limit. In particular, $h$ annihilates the TFD state \eqref{eq:TFD}. In the bulk dual, this corresponds to an isometry of the bulk solution under evolution by the bulk dual of $h$. This isometry also becomes important when defining a geometric phase for time translations for the black hole as in \cite{Nogueira:2021ngh}, whose existence is in fact intimately related to the non-trivial centre of the algebra. Before discussing this in detail in the next section, we will first complete reviewing the algebras acting on the Hilbert space of the eternal black hole by discussing $1/N$ corrections \cite{Witten:2021unn}.

The above discussion is valid in the large $N$ limit. However, taking into account $1/N$ corrections, the right hand side of \eqref{eq:CentralOperator} no longer vanishes. Therefore, $U$ does not define a central operator. To properly include it into the algebra, it was observed that the previous definition of $U$ also receives $1/N$ corrections. These corrections resemble that $[U,\mathcal{O}]\neq0$ when including terms proportional to $1/N$. The resulting algebra is then defined by taking the crossed product of ${\cal A}_{L,0}$ with the algebra of bounded functions of $\hat{h}+X$, where $X=\beta NU$ and $\hat{h}$ corresponds to the bulk dual of $h$. The crossed product in ${\cal A}_L={\cal A}_{L,0}\rtimes{\cal A}_{\hat{h}+X}$ indicates that the two pieces composing ${\cal A}_L$ do not commute.

As shown in \cite{Witten:2021unn}, ${\cal A}_{\hat{h}+X}$ can be understood as the modular automorphism group of the type III$_1$ algebra ${\cal A}_{L,0}$. Taking the crossed product of a type III$_1$ algebra with its modular automorphism group yields a factor of type II$_\infty$ \cite{Takesaki:1973duality,Connes:1973classification,Connes:1994yd}. Since the resulting algebra after taking the crossed product is of type II, it allows for defining a trace. Moreover, since it is a factor, the centre is now trivial.

\subsubsection{Topology of Phase Space and Centre of the Algebra}
\label{sec:TopPhaseSpaceCentre}

In the preceding section we have briefly reviewed which operator algebras are found for the eternal black hole. In particular, there is a transition between a type III algebra, present in the large $N$ limit, and a type II algebra, obtained by including $1/N$ corrections to the large $N$ limit. In the following we discuss how this change of algebra is related to the geometric phase defined for the AdS eternal black hole in \cite{Nogueira:2021ngh}.

In the eternal black hole spacetime, there is no globally defined time-like Killing vector. This is due to the fact that a time-like Killing vector defined locally switches sign when passing the horizon \cite{Verlinde:2020upt}. Therefore, time is not defined globally for the eternal black hole. Rather, there are two time coordinates $t_L$ and $t_R$ associated to each of the boundaries. A natural choice for the relation between these times at the boundary is to simply identify them, $t_L=t_R$. However, since time is not defined globally, this identification is not guaranteed to hold deep in the bulk. In particular, at the horizon a different identification may be imposed, $t_L=2\delta-t_R$, where $\delta$ is the relative offset between the left and right times. The offset $\delta$ is locally invisible. Its value can only be determined by a non-local measurement. However, inserting $t_L=t_R$ into $t_L=2\delta-t_R$ shows that $\delta=t_L=t_R$ can be used to describe time evolution on both boundaries.

In fact, $\delta$ corresponds to a degree of freedom in the bulk. This can be seen as follows. As mentioned in the introduction, the bulk Hilbert space may be obtained by defining a fibre bundle over the space of classical solutions ${\cal G}_M$, using insights from AdS/CFT.  The Hilbert space is found by analysing the asymptotic symmetries of the bulk spacetime, described by groups $G_L$ and $G_R$. These groups typically include time translations as well as rotations. The full asymptotic symmetry group then follows as $G_L\times G_R$. To obtain ${\cal G}_M$, this product has to be quotiented by the isometries of the bulk spacetime. The isometries can be understood as trivial bulk diffeomorphisms, i.e. diffeomorphisms that do not change the state in the CFT. In the CFT picture, isometries are reflected by the fact that the CFT state is annihilated by the difference of the corresponding asymptotic charges $Q_L$ and $Q_R$. We already mentioned above that the TFD state is annihilated by the difference of the boundary Hamiltonians $h$. This corresponds to the bulk isometry for trivial time translations. The full isometry group, which is the diagonal subgroup of $G_L\times G_R$, is denoted by $G_D$ and the moduli space of classical bulk solutions is given by
\begin{align}
    {\cal G}_M=\frac{G_L\times G_R}{G_D}.\label{eq:ModuliSpace}
\end{align}
This space contains parameters $g$ that fix the particular bulk solution. As such, it contains parameters determining the relative alignment between the left and right coordinates, in particular also the time coordinates. So $\delta$ is a bulk degree of freedom.

\begin{figure}[t]
    \centering
    \includegraphics{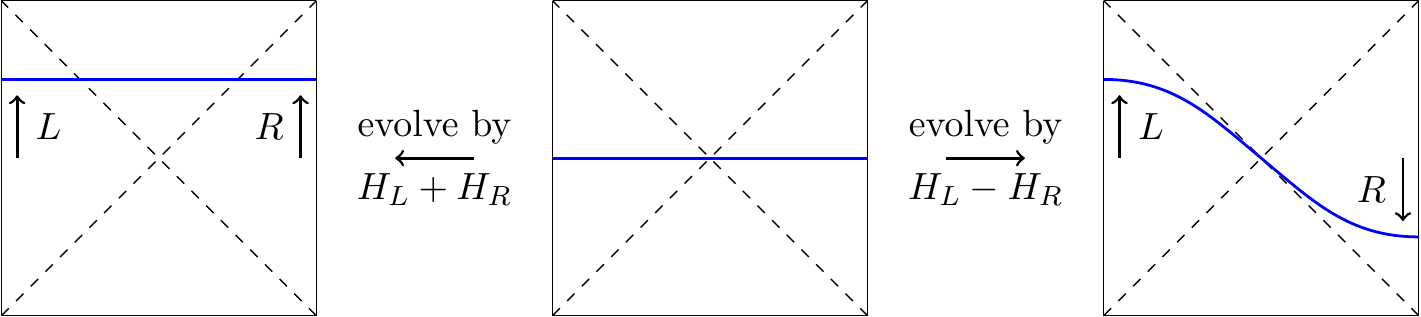}
    \caption{Visualisation of different time evolutions of the TFD state. In the central panel, the holographic dual of the usual TFD state \eqref{eq:TFD} at the $\delta=0$ slice is depicted. The blue line represents the slice at which \eqref{eq:TFD} is defined. The left panel shows the holographic dual to the TFD state time-evolved by $H_L+H_R$, which changes the state in the CFT. The panel on the right shows the time evolution by $H_L-H_R$, which leaves the CFT state invariant.}
    \label{fig:TimeEvolvedTFD}
\end{figure}

In the TFD state \eqref{eq:TFD}, the slice $\delta=0$ is chosen. Choosing different slices with $\delta\neq0$ leads to time-shifted TFD states, obtained by evolving \eqref{eq:TFD} using $H_L+H_R$ for a time period of $\delta$. This is visualised in fig.~\ref{fig:TimeEvolvedTFD}. The entanglement properties of the original TFD state are unaffected by this evolution. The time-shifted TFD states are interpreted as microstates of the black hole \cite{Papadodimas:2015jra} (see also \cite{Banerjee:2023liw} for a discussion of the time-shifted TFD states as black hole microstates in the sense of \cite{Balasubramanian:2022gmo,Balasubramanian:2022lnw}). Correspondingly, the action of $H_L+H_R$ is non-trivial in the phase space and thereby alters the CFT state, as opposed to the action of $H_L-H_R$. This is reflected by the fact that in the phase space, the variable $\delta$ appears with a dual variable $H=H_L+H_R$. In contrast, $h$ is not part of the phase space; see e.g. \cite{Harlow:2018tqv} for an explicit analysis in two-dimensional Jackiw--Teitelboim (JT) gravity.

Time translations are always part of the asymptotic symmetry group. Since these are generated by the Hamiltonian, which is the operator of importance in discussing the transition between the algebra types III$_1$ and II$_\infty$, we focus on time translations from now on. Time translations contribute to $G_L$ and $G_R$ with a factor of $\U(1)$ each. The isometry of the bulk is described by time evolution with $h$ and is therefore also described by $\U(1)$. Therefore, the corresponding part of the moduli space of classical solutions is given by
\begin{align}
    \frac{\U(1)\times\U(1)}{\U(1)}\sim\U(1)\sim S^1.\label{eq:ModSpaceTimeTrans}
\end{align}
The circle is parametrised by $\delta$. Since $S^1$ is not simply connected,\footnote{A simple way to see this is to consider a path winding twice around $S^1$. This path cannot be deformed smoothly, i.e. without leaving the manifold, into a path winding around $S^1$ only once.} the fibre bundle defined above $S^1$ is non-trivial. The corresponding geometric phases are related to winding numbers. To see this explicitly, following \cite{Nogueira:2021ngh}, consider time evolution of the TFD state by $H=H_L+H_R$ by an amount $\delta$. The connection is then defined as
\begin{align}
    A_\delta=\i\langle\tfd|U^\dagger\partial_\delta U|\tfd\rangle,\label{eq:ConnectionTFD}
\end{align}
where $U=e^{\i(H_L+H_R)\delta}$. From this time evolution also follows that $\delta$ are periodic as $\delta\sim\delta+\frac{\pi}{E_n}$ for each individual value of $E_n$. Collecting all of these circles together yields the punctured plane $\mathds{R}^2\backslash\{0\}$. Closed paths encircling the puncture are not contractible but pick up winding numbers.

Let us illustrate this by the two-spin example discussed in sec.~\ref{sec:TwoSpins}. We have derived the TFD state of this system in sec.~\ref{sec:EntanglementTemperature} as
\begin{align}
    |\psi_0\rangle=\frac{1}{\sqrt{1+e^{-\beta_{\text{ent}}2\mu_BB}}}\Big[|\!\downarrow\!\tilde{\downarrow}\rangle+e^{-\beta_{\text{ent}}\mu_BB}|\!\uparrow\!\tilde{\uparrow}\rangle\Big].
\end{align}
Since this state provides a thermal description of the entanglement between the two spins, the natural Hamiltonians to use for the evolution are the modular Hamiltonians of the first and second spin,
\begin{align}
    H_{\text{mod},1/2}=\frac{1}{2}\ln\frac{2\pi+\Phi_G}{2\pi-\Phi_G}\,\sigma_z=\beta_{\text{ent}}\mu_BB\,\sigma_z.\label{eq:TwoSpinModHam}
\end{align}
Note that in \eqref{eq:ConnectionTFD}, the evolution is performed with the physical Hamiltonians $H_{L/R}$. Since the usual TFD state describes a thermal system, the modular Hamiltonian in this case is given by $H_{\text{mod},L/R}=\beta H_{L/R}$. While the connection $A_\delta$ defined in \eqref{eq:ConnectionTFD} and the periodicity of $\delta$ will change accordingly under this rescaling, the resulting geometric phase after integrating the connection is invariant.

Evolving $|\psi_0\rangle$ by $H_{\text{mod},1}-H_{\text{mod},2}$ leaves the state invariant, resembling the invariance of the TFD state \eqref{eq:TFD} under evolution of $H_L-H_R$. On the other hand, evolution by the sum of the modular Hamiltonians for some duration of time $\delta$ (modular time in this example)  gives rise to the state
\begin{align}
    U|\psi_0\rangle&=e^{\i(H_{\text{mod},1}+H_{\text{mod},2})\delta}|\psi_0\rangle\notag\\
    &=\frac{1}{\sqrt{1+e^{-\beta_{\text{ent}}2\mu_BB}}}\Big[e^{-\i\beta_{\text{ent}}2\mu_BB\delta}|\!\downarrow\!\tilde{\downarrow}\rangle+e^{\i\beta_{\text{ent}}2\mu_BB\delta}e^{-\beta_{\text{ent}}\mu_BB}|\!\uparrow\!\tilde{\uparrow}\rangle\Big].\label{eq:TimeEvolvedTwoSpinState}
\end{align}
As we mentioned below \eqref{eq:ConnectionTFD}, time evolution by the sum of the Hamiltonians leads  to a periodicity $\delta_P$. Evolving for this amount of time results in exactly the same state as the initial state. With the above result \eqref{eq:TimeEvolvedTwoSpinState} we find that this periodicity is given by
\begin{align}
    \delta_P=\frac{\pi r}{\beta_{\text{ent}}\mu_BB},
\end{align}
where $r\in\mathds{Z}$ is an integer.

Evaluating \eqref{eq:ConnectionTFD} for the time evolved TFD state of the two-spin system \eqref{eq:TimeEvolvedTwoSpinState}, the connection is given by
\begin{align}
    A_\delta=2\beta_{\text{ent}}\mu_BB\frac{1-e^{-\beta_{\text{ent}}2\mu_BB}}{1+e^{-\beta_{\text{ent}}2\mu_BB}}.
\end{align}
Integrating over $\delta$ results in the geometric phase
\begin{align}
    \Phi_G^{(\text{TFD})}=\int_0^{\delta_P}\d\delta\,A_\delta=2\pi r\frac{1-e^{-\beta_{\text{ent}}2\mu_BB}}{1+e^{-\beta_{\text{ent}}2\mu_BB}}=\Phi_Gr,\label{eq:TopPhaseTwoSpin}
\end{align}
where $\Phi_G$ is the geometric phase calculated as the symplectic volume in \eqref{eq:SympVolume}. As we pointed out earlier below \eqref{eq:TwoSpinModHam}, the same result for $\Phi_G^{(\tfd)}$ is obtained when the evolution of $|\psi_0\rangle$ is performed with the sum of the rescaled modular Hamiltonians $\tilde{H}_{\text{mod},1/2}=H_{\text{mod},1/2}/\beta_{\text{ent}}$.

The integer $r$ in \eqref{eq:TopPhaseTwoSpin} counts how many times the time evolution wraps around $S^1$. In terms of the evolution, this may also be phrased as $r$ counting how many times the time evolved state returns to the initial state. Interestingly, while usually the winding number comes with a prefactor of $2\pi$, the prefactor of $r$ in \eqref{eq:TopPhaseTwoSpin} is given by the geometric phase of the two-spin system. This however matches perfectly with the fact that $\Phi_G$ is defined as the symplectic volume. While in the usual definition, the factor $2\pi$ is the volume of $S^1$, this changes for non-vanishing entanglement between the spins. The circle, while still possessing the topological features of $S^1$, is parametrised by an angular variable with a periodicity different from $2\pi$ and accordingly, its volume changes. In particular, we find that the geometric phase \eqref{eq:TopPhaseTwoSpin} always vanishes for maximal entanglement. This can be understood by the fact that for maximal entanglement, time evolution by the sum of the modular Hamiltonians leaves the state completely invariant, as happens for evolution by the difference of the modular Hamiltonians already for arbitrary entanglement.

The transition between the type III and type II algebras in the holographic setting of the eternal black hole can be understood using the spin example discussed above. In the large $N$ limit the eternal black hole admits a description in terms of an algebra of type III$_1$ \cite{Leutheusser:2021frk,Leutheusser:2021qhd}, as reviewed in sec.~\ref{sec:AlgebraicSetupEternalBlackHole}. This algebra does not have a trace. As discussed in sec.~\ref{sec:TraceTypeIITypeIII} for the spin system, this can be understood by the fact that none of the states on which an algebra of type III acts is maximally entangled, or equivalently, by the presence of the geometric phases $\Phi_G^{(n)}$ in the states \eqref{eq:EntangledSpinGeomPhase}. Since the TFD state is understood as the dual description of an eternal black hole, with the microstates given by the time-shifted TFD states \cite{Papadodimas:2015jra}, our discussion of the spin system very naturally generalises to the algebraic description of black holes. Using the spin microstates of the TFD state, we already demonstrated that the geometric phases $\Phi_G^{(\text{TFD})}$ computed in \eqref{eq:TopPhaseTwoSpin} can only be non-trivial if the geometric phases $\Phi_G^{(n)}$ do not vanish. Therefore, the geometric phase associated to time translations \cite{Nogueira:2021ngh} can be defined for every state of a type III algebra. In the case where all of the geometric phases $\Phi_G^{(n)}$ vanish, the state is maximally entangled. It allows to define a trace and therefore corresponds to an algebra of type II. For this particular state, which is the cyclic and separating vector of the type II algebra, all of the geometric phases $\Phi_G^{(\text{TFD})}$ computed in \eqref{eq:TopPhaseTwoSpin} vanish. Although this construction uses spin microstates, it is generically applicable to any set of microstates giving rise to the TFD state \cite{Cottrell:2018ash}. Therefore this construction naturally encompasses the algebra of operators dual to a black hole spacetime. To summarise, the geometric phase associated to time translations \cite{Nogueira:2021ngh} is defined for every state of a type III algebra, but for a type II algebra, there exists one particular state where all of the geometric phases vanish. It is this particular state which consistently defines the trace for the type II algebra, as shown by our earlier result \eqref{eq:CheckCyclicityTypeIII}. We have therefore found an explicit realisation of the relation between geometric phases and the trace as discussed in sec.~\ref{sec:TraceTypeIITypeIII}, for the AdS eternal black hole.

Finally, we point out that the geometric phase is related to the non-trivial common centre of the type III$_1$ algebras of \cite{Leutheusser:2021frk,Leutheusser:2021qhd}. The geometric phase computed by integrating the connection \eqref{eq:ConnectionTFD} is associated to time translations, generated by the boundary Hamiltonians \cite{Nogueira:2021ngh}. The boundary Hamiltonians are both related to the central operator $U$ \cite{Leutheusser:2021frk,Leutheusser:2021qhd}, as reviewed in sec.~\ref{sec:AlgebraicSetupEternalBlackHole}, which is a shared degree of freedom. This collective coordinate shared by the left and right algebras is given by the mass of the black hole. This is made explicit by \cite{Henneaux:2019sjx,Banerjee:2022jnv} where the mass of the black hole shows up as a coupling term in the chiral boson action, defined on an annulus geometry. Due to this shared mode, the left and right type III algebras share a common centre. In the absence of this shared mode, $H_L$ and $H_R$ correspond to different bulk isometries, each represented by $\U(1)$. Accordingly, the moduli space \eqref{eq:ModSpaceTimeTrans} will be topologically trivial. In this case, there is no ambiguity in relating the left and right boundary times. Following this argument, the non-vanishing of the geometric phase associated to time translations \cite{Nogueira:2021ngh} may therefore be interpreted as a signal for non-factorisation of the operator algebras, in that the algebras have a non-trivial common centre. Once $1/N$ corrections are included, the centre becomes trivial and the algebra is deformed to type II factors \cite{Witten:2021unn}. In this case, as discussed in sec.~\ref{sec:TraceTypeIITypeIII}, there exists a state where the geometric phase of \cite{Nogueira:2021ngh} vanishes. In particular, this is the cyclic and separating vector of the type II algebra.

\section{Geometric Phase and Missing Information}
\label{sec:GeometricPhaseAndMissingInformation}

The temperature of a black hole is a consequence of the information hidden behind the horizon. In sec.~\ref{sec:EntanglementTemperature} we found that the entanglement temperature is determined by the geometric phase. We therefore expect that a non-vanishing geometric phase can also be interpreted as indicating missing information. In the following, we elaborate on this relation by showing that geometric phases are a signature of missing information about the microscopic structure of the phase space. The missing information has its origin in the inability of a local observer to access the full Hilbert space and is present both in systems with and without entanglement. In systems without entanglement, the missing information is related to the existence of a global symmetry. The global symmetry allows to distinguish between the projective Hilbert space and the full Hilbert space of the system. A local observer only has access to the former. In a system with entanglement between two subregions, the local observer located in one of the subregions cannot access information arising from global symmetries of the full system. The global symmetry generates additional relative phases in the state describing the full system. These relative phases do not impact measurements of an observer in a subregion as they leave invariant the density matrix of the subregion.

We begin in sec.~\ref{sec:Examples without entanglement} by discussing the relation between geometric phases and missing information in systems without entanglement in two examples, namely the well-known spin in a magnetic field \cite{Berry:1984jv} and Virasoro Berry phases \cite{Oblak:2017ect} in a CFT. In sec.~\ref{sec:EntBerryPhasesInQFT}, we then discuss missing information for Berry phases in two entangled CFTs dual to the eternal AdS black hole and modular Berry phases for subregions in CFTs.

\subsection{Examples without Entanglement}
\label{sec:Examples without entanglement}

We now discuss the relation between missing information and Berry phases for the examples of a single spin in a magnetic field in sec.~\ref{sec:SingleSpin} and for a two-dimensional CFT in sec.~\ref{sec:Virasoro Berry phase}.

\subsubsection{Single Spin System}
\label{sec:SingleSpin}

The relation between missing information and the geometric phase is already present in systems without entanglement. To illustrate this, we use the example of a single spin coupled to a magnetic field. As we will see in more detail in the following, the missing information is a consequence the necessity to use (at least) two coordinate patches to fully cover the projective Hilbert space. The Hamiltonian of the system is given by
\begin{align}
    H=J\vec{B}\cdot\vec{S}=\frac{JB}{2}\vec{n}\cdot\vec{\sigma},
\end{align}
where $\vec{B}=B\vec{n}$ with $\vec{n}$ a radial unit vector and $\vec{S}=\frac{1}{2}\vec{\sigma}$. The unit vector $\vec{n}$ depends on the two angular coordinates $\phi\in[0,2\pi)$ and $\theta\in[0,\pi]$ parameterising the Bloch sphere. The projective Hilbert space is given by $\CP{1}\sim S^2$. Global phases to the eigenvectors of $H$ then constitute a $\U(1)$ bundle over the projective Hilbert space $\CP{1}$. This is essentially the Hopf fibration \cite{hopf1964abbildungen} mentioned in the introduction as an example for a non-trivial bundle.

To see explicitly that this bundle is non-trivial, the ground state of the above Hamiltonian is given by
\begin{align}
    |\tau\rangle&=-e^{-\i\phi}\sin\frac{\theta}{2}|\!\uparrow\rangle+\cos\frac{\theta}{2}|\!\downarrow\rangle,
\end{align}
where we assumed w.l.o.g. that $JB>0$. This state becomes singular at the north pole of $S^2$ where $\theta=0$ since $\phi$ is not defined there. Since an overall phase does not change the physical properties of the state, we may use the state $|\tilde\tau\rangle$ obtained by multiplying $|\tau\rangle$ with $e^{\i\phi}$,
\begin{align}
    |\tilde\tau\rangle&=-\sin\frac{\theta}{2}|\!\uparrow\rangle+e^{\i\phi}\cos\frac{\theta}{2}|\!\downarrow\rangle.
\end{align}
This state is non-singular at $\theta=0$. However, a singular point is still present and now located at the south pole of $S^2$ where $\theta=\pi$.\footnote{Had we assumed that $JB<0$, the corresponding ground state $|\tau^\p\rangle$ would be singular at $\theta=\pi$ and the corresponding transformed ground state $|\tilde{\tau}^\p\rangle$ at $\theta=0$. In this version, the ground state $|\tau^\p\rangle$ defines the north pole patch while $|\tilde{\tau}^\p\rangle$ defines the south pole patch. This does not change the result for the geometric phase.} Since there are no other singular points, the combination of the two coordinate patches where the states $|\tau\rangle$ and $|\tilde\tau\rangle$ are defined fully covers the projective Hilbert space $\CP{1}$. For each of the patches, typically referred to as north pole and south pole patches, a connection can be defined using the corresponding state, $|\tilde\tau\rangle$ and $|\tau\rangle$ respectively,
\begin{align}
    A^{(N)}&=\i\langle\tilde\tau|\d|\tilde\tau\rangle=\frac{-1-\cos\theta}{2}\d\phi,\\
    A^{(S)}&=\i\langle\tau|\d|\tau\rangle=\frac{1-\cos\theta}{2}\d\phi.
\end{align}
Note that at the points where $|\tau\rangle$ and $|\tilde{\tau}\rangle$ are singular, the corresponding connections, $A^{(S)}$ and $A^{(N)}$ respectively, vanish. The connections are related by the aforementioned $\U(1)$ transformation $U=e^{\i\phi}$ as
\begin{align}
    A^{(S)}&=A^{(N)}-\i U^\dagger\d U.
\end{align}
The curvature of the connections is given by
\begin{align}
    F=\d A^{(N)}=\d A^{(S)}=\frac{1}{2}\sin\theta\,\d\theta\wedge\d\phi.\label{eq:BerryCurvature}
\end{align}
The geometric phase is found by integrating $F$ over the full projective Hilbert space $\CP{1}\sim S^2$. The result does not vanish, showing that the fibre bundle is non-trivial. As we saw in the above derivation, this is tied to the necessity to use two patches in order to fully cover the projective Hilbert space.

The geometric phase contains information about the geometry and topology of the projective Hilbert space. In particular, the curvature allows to compute its Euler characteristic $\chi$ by use of the Chern theorem,
\begin{align}
    \chi({\cal M})=\int_{\cal M}e(F)=\frac{1}{(2\pi)^n}\int_{\cal M}\sqrt{\det(F)},
\end{align}
given here for a general (suitable) manifold ${\cal M}$ of $2n$ real dimensions. Also, the Euler characteristic is related to the number of holes $g$ of a manifold as $\chi=2-2g$. Inserting \eqref{eq:BerryCurvature} and using that in our case $n=1$ yields the well-known result
\begin{align}
    \chi(S^2)=2\quad\text{and correspondingly,}\quad g=0.
\end{align}
We see that knowledge of the curvature of the bundle allows to calculate that the base space (in this case) is topologically trivial as it has no holes. Every two-dimensional manifold without holes is topologically equivalent to $S^2$, which as mentioned earlier is indeed the geometry of the base space. Such topological quantities as the Euler characteristic are not only present in the field of high energy physics, but also arise in condensed matter physics. The most famous example is the quantum Hall effect, where the Hall conductance is quantised in terms of Chern numbers \cite{Thouless1982quantized,Hastings2015quantization}.

This discussion demonstrates how the information provided by the geometric phase determines a simple single-spin system. In the next section we will turn to the  more advanced example of Virasoro Berry phases in  CFTs.

\subsubsection{Virasoro Berry Phase in a Single Boundary Geometry}
\label{sec:Virasoro Berry phase}

Here we discuss missing information in systems without entanglement for the example of a geometric phase in a two-dimensional CFT. We show that the missing information that gives rise to the geometric phase is related to the existence of a global symmetry in the CFT.

The Virasoro Berry phase in a CFT dual to a spacetime with a single boundary was first derived in \cite{Oblak:2017ect}. It arises from applying local conformal symmetry transformations to a highest-weight state of the CFT in the presence of global symmetries. In two dimensions, the conformal transformations are the diffeomorphisms of the unit circle $f\in\text{Diff}(S^1)$. The Virasoro group is the central extension of $\text{Diff}(S^1)$, which we denote by $\widehat{\text{Diff}}(S^1)$. Due to the central extension, group elements are pairs $(f,\alpha)\in\widehat{\text{Diff}}(S^1)$, where $\alpha\in\mathds R$ is central. Irreducible representations are formed by Verma modules with highest-weight states $\ket{h}$. Depending on the value of the conformal weight $h$, the state has different global symmetries. For $h=0$, which corresponds to a CFT in the vacuum, the symmetry is $\text{SL}(2,\mathds{R})$, whereas for $h>0$, it is $\U(1)$. In the following we focus on the case $h>0$.

The $\U(1)$ symmetry transformations are generated by the CFT Hamiltonian and imply time translation invariance of the CFT. Taking into account the left- and right-moving sector of the CFT, the Hamiltonian reads $H=L_0+\bar{L}_0$. Acting with the $\U(1)$ subgroup of the conformal group on the state $\ket{h}$ only yields an overall phase to the state, $\ket{h}\rightarrow e^{\i\gamma}\ket{h}$ which cannot be measured. The state $\ket{h}$ belongs to the projective Hilbert space and represents all physically equivalent states $e^{\i\gamma}\ket{h}$ in the full Hilbert space of the system. A local observer only has access to the projective Hilbert space, but not to information related to the global symmetry. Therefore, the global symmetry group $\U(1)$ represents the fibre of the fibre bundle in fig.~\ref{fig:BaseManifoldFibreAndHolonomy}. On the other hand, the base space of the fibre bundle is then formed by the set of transformations that physically change the state $\ket{h}$ and are given by $\frac{\widehat{\text{Diff}}(S^1)}{\U(1)}$. Upon quantisation, this gives rise to the projective Hilbert space.

Moving along a closed path in the base space $\frac{\widehat{\text{Diff}}(S^1)}{\U(1)}$, the state considered transforms non-trivially under the conformal transformation $f$.  At each point along the path, it has a phase ambiguity due to the global symmetry $\U(1)$. This phase ambiguity may be interpreted as a freedom to choose an origin for the time coordinate. Since all expectation values are invariant, the phase cannot be measured by an observer and represents missing information. An observer can therefore only distinguish states in the projective Hilbert space and loses information regarding the state in the full Hilbert space. In particular, the observer only has access to the overall phase accumulated when they return to the original state (up to the phase). This signals the missing information. The Berry phase is obtained by integrating the non-vanishing curvature form on the manifold $\frac{\widehat{\text{Diff}}(S^1)}{\U(1)}$ over the surface enclosed by the closed path through the manifold. The curvature form is given by the Kirillov-Kostant symplectic form on the coadjoint orbit of the Virasoro group and yields the Berry phase \cite{Alekseev:1988ce,Oblak:2017ect}
\begin{equation}
   \Phi_B=S_{\mathrm{geo}}^{\pm}\left[f,b_0\right]=\int\d t\d\sigma\left(b_0 f^{\prime} \partial_{\pm}f+\frac{c}{12} \frac{f^{\prime \prime} \partial_{\pm} f^{\prime}}{\left(f^{\prime}\right)^2}\right)\label{eq:Virasoro Berry phase}
\end{equation}
up to a boundary term. Here, $b_0$ denotes the expectation value of the energy-momentum tensor of the CFT in the state $\ket{h}$, $b_0=\frac{1}{2\pi}\bra{h}T\ket{h}$.

To summarise, the Virasoro Berry phase arises due to the existence of a global symmetry in the CFT combined with a geometrically non-trivial base space. The existence of the global symmetry represents missing information about the choice of time coordinate for the global state at each point along the closed path in the base space.

\subsection{Examples with Entanglement}
\label{sec:EntBerryPhasesInQFT}

We now move on to discuss how the Berry phase signals missing information in systems with entanglement between subregions. Here, the missing information is related to the existence of relative phases in the global state describing the full system. This cannot be accessed by an observer located in a subregion. The relative phases exists due to independent `global' symmetries in each subregion. We discuss the missing information for the examples of the Virasoro Berry phase and the gauge Berry phase, both in the presence of a spacetime wormhole, in sec.~\ref{sec:Virasoro Berry phase in the presence of a wormhole} and sec.~\ref{sec:GaugeBerryPhase}. We also consider modular Berry phases in a two-dimensional CFT in sec.~\ref{sec:Modular Berry phases}.

\subsubsection{Virasoro Berry Phase in the Presence of a Wormhole}
\label{sec:Virasoro Berry phase in the presence of a wormhole}

Here we consider the Virasoro Berry phases of sec.~\ref{sec:Virasoro Berry phase} for CFTs dual to an eternal AdS black hole. We show that the missing information originates from an independent choice of time coordinate in each CFT that gives rise to a relative phase in the TFD state of the full system. 

The eternal AdS black hole implies the presence of a wormhole in the bulk. Moreover, there are now two asymptotic regions, i.e.~also two boundaries, on each of which a copy of the CFT is defined, see. fig.~\ref{fig:EternalBlackHole}. Since both CFTs are causally separated by the horizon, we may independently apply conformal transformations on each boundary, $\{f_R,f_L\}\in \text{Diff}(S^1)$. Additionally, there are two copies of the Hamiltonian, $H_L$ and $H_R$, giving rise to the global asymptotic symmetry group $\U(1)\times\U(1)$. The Hamiltonians generate time translations on the left and right boundaries. This induces an independent choice of time coordinate for each boundary. From this perspective, we may naively expect that we obtain two copies of the single CFT Berry phase \eqref{eq:Virasoro Berry phase} corresponding to fibre bundles with base space $\frac{\widehat{\text{Diff}}(S^1)}{\U(1)}\times \frac{\widehat{\text{Diff}}(S^1)}{\U(1)} $. This is not true, however. As shown in \cite{Henneaux:2019sjx,Banerjee:2022jnv}, there are additional constraints due to the presence of the wormhole in the bulk. In particular, the black hole mass or equivalently its energy must be the same when measured from either boundary. This is reflected in the invariance of the TFD state under the combined action of $H_L-H_R$ as explained in sec.~\ref{sec:TopPhaseSpaceCentre}. For the Berry phase, the existence of the additional constraint implies that the value of $b_0$ in \eqref{eq:Virasoro Berry phase} must be the same both for the left and the right CFT since $b_0$ is related to the black hole mass by $b_0=\frac{M}{32\pi G_N}$ \cite{Henneaux:2019sjx}, in the case of a non-rotating black hole. Therefore, the mass acts as a coupling between Berry phases in the left and right boundary. Up to a boundary term, the Berry phase in the presence of a wormhole then reads
\begin{align}
    \Phi_B&=S_{\text{geo}}^{-}\left[f_L,b_0\right]-S_{\text{geo}}^{+}\left[f_R,b_0\right].\label{eq:coupled_Berry_phase}
\end{align}
The Virasoro Berry phase in the presence of a wormhole originates from independent misaligned choices of the time coordinate for the left and right CFTs. An observer on the left and right boundary may choose an independent origin for their time coordinate, as they cannot communicate with each other. The missing information for an observer is therefore the overall misalignment of the time frame, which cannot be measured by either observer. The misalignment is generated by a $\U(1)$ transformation and gives rise to the fibre bundle structure with base space $\frac{\widehat{\text{Diff}}(S^1)\times\widehat{\text{Diff}}(S^1)}{\U(1)} $ and fibre $\U(1)$.

Let us now discuss this result in the context of vN algebras. As reviewed in sec.~\ref{sec:AlgebraicSetupEternalBlackHole}, in the large $N$ limit, holographic CFTs  are type III vN algebras with a dual effective bulk field theory in the limit $G_N\rightarrow 0$. The operator algebras of the CFTs on the left and right boundaries are given by ${\cal A}_{L/R}={\cal A}_{L/R,0}\otimes{\cal A}_U$. The central operator $U$ is given by $U=\frac{H_{L/R}}{N}$ \cite{Witten:2021unn}.
The centre is related to the mass of the black hole, or equivalently the common $b_0\propto M$ in \eqref{eq:coupled_Berry_phase}. In particular, since the black hole mass is the same when measured from either boundary, the centre is common to both CFTs. In terms of the operator algebras, this is reproduced by including the same ${\cal A}_U$ both in ${\cal A}_L$ and ${\cal A}_R$, so the centre is shared by the two algebras. In the language of Virasoro Berry phases, this is reflected in the structure of the base space $\frac{\widehat{\text{Diff}}(S^1)\times\widehat{\text{Diff}}(S^1)}{\U(1)}$ by the fact that the quotient is taken only by a single $\U(1)$ group.

\subsubsection{Gauge Berry Phase}
\label{sec:GaugeBerryPhase}

Here we consider the modular time evolution in the presence of an eternal AdS black hole. Since both exterior regions are separated by a horizon, observers on the left and right boundary may independently choose their modular time coordinates $s_L$ and $s_R$. If their modular times are aligned, the TFD state is given by
\begin{equation}
    |\text{TFD}\rangle=\frac{1}{\sqrt{Z}}\sum_Ee^{-\beta\frac{E}{2}}|E_L\rangle|E_R^\ast\rangle.
    \label{eq:modular_TFD}
\end{equation}
Note that here, $|E_{L/R}\rangle$ is the eigenbasis of the modular Hamiltonian, while in \eqref{eq:TFD}, $|n_{L/R}\rangle$ is the eigenbasis of the physical Hamiltonian. If the modular times of the observers are not aligned, the global state differs by a relative phase from the state \eqref{eq:modular_TFD} and reads
\begin{equation}
    |\text{TFD}\rangle_{\delta'}=\frac{1}{\sqrt{Z}}\sum_Ee^{\i 2E\delta^\p}e^{-\beta\frac{E}{2}}|E_L\rangle|E_R^\ast\rangle,\label{eq:modular_TFD_phase}
\end{equation}
where $2\delta'=s_L+s_R$ is a relative difference in the left and right modular time. This may be understood as follows. The horizon separates the left and right exterior regions, and each subregion is described by a reduced density matrix which is not sensitive to the relative difference $\delta^\p$. Therefore, from the perspective of an observer on the left boundary, the information about the relative phase in the global state of the CFT is missing. The observer cannot access the choice of modular time on the right boundary. Following the approach of sec.~\ref{sec:TopPhaseSpaceCentre}, we may then define a Berry phase for which the modular time shift $\delta'$ is $\frac{\pi}{E}$-periodic with Berry connection
\begin{equation}
    A_{\delta'}=\i\,_{\delta^\p}\!\langle\tfd|\partial_{\delta^\p}|\tfd\rangle_{\delta^\p}.\label{eq:Berry_phase_delta}
\end{equation}
The Berry phase signals missing information about the structure of the Hilbert space from the point of view of an observer located in the left exterior. An observer in the left exterior measures expectation values of observables with respect to the density matrix $\rho_L$, as they do not have access to the global state. Observers in the left and right exterior regions may then independently choose the origins of their modular time. This leads to a misalignment of the modular time coordinates in the subregions and yields the relative phase in the global state \eqref{eq:modular_TFD_phase}. The Berry phase associated to the modular Berry curvature $A_{\delta'}$ is topological since the periodicity of $2\delta'=s_L+s_R$ leads to a non-trivial topology $\mathds{R}^2\backslash\{0\}$ of the base space. As explained in sec.~\ref{sec:TopPhaseSpaceCentre} such an entanglement Berry phase is in relation to the non-existence of a trace on a type III algebra and is absent for a particular state by transitioning to a type II vN algebra.

Moreover as we elaborated in sec.~\ref{sec:AlgebraicPerspective} around \eqref{eq:TopPhaseTwoSpin}, this geometric phase has an interpretation in terms of winding numbers due to the non-trivial topology, as discussed above. A concrete example of this was found in the context of JT gravity \cite{Nogueira:2021ngh}. Let us now explain how these winding numbers are related to missing information and the operator algebra. Within geometric quantisation, winding numbers appear in the spectrum of the `prequantum' momentum operator (see e.g.~\cite{carosso2018geometric}). In JT gravity, this operator corresponds to the Hamiltonian. In general, the spectrum of this operator takes the form $\{r+\lambda,r\in\mathds{Z}\}$, where $r$ is the winding number and $\lambda\in[0,1)$ is an ambiguity parameter that leaves the symplectic form invariant. However, since $\lambda$ appears in the spectrum, different values of $\lambda$ correspond to inequivalent prequantum operators. Thus, the algebra of observables is sensitive to this ambiguity parameter $\lambda$. Given a fixed $\lambda$, we may define an equivalence of prequantum momentum operators corresponding to different values of $r$. All these operators correspond to the same symplectic form and spectrum and therefore are physically indistinguishable for a local observer. The information about $r$ can therefore be termed  missing information, since we cannot distinguish different elements within the same equivalence class. In the context of AdS/CFT, this information corresponds to the specific gluing of the bulk and boundary spacetimes (cf.~the discussion around \eqref{eq:ModuliSpace}).

\subsubsection{Modular Berry Phase in the Presence of a Wormhole}
\label{sec:Modular Berry phases}

 Similarly to the Virasoro Berry phases in the presence of a wormhole, modular Berry phases arise from symmetries in the subregions of a system. In contrast to the Virasoro Berry phase, these symmetries are not generated by the physical Hamiltonian but by the modular Hamiltonian. As we will see, in this case the missing information is a relative phase in the global state describing the full system which cannot be measured by an observer restricted to a subregion. The relative phase originates from a misalignment of the modular time parameters in each subregion. This may be understood as follows.

The modular Hamiltonian is formally defined as  $H_{\mathrm{mod},A}=-\ln(\rho_A)$ and generates an automorphism that maps operators $\mathcal{O}$ of the algebra $\mathcal{A}_{A}$ to themselves,
\begin{equation}
	U(s) \mathcal{O} U(-s)=\mathcal{O} , \text{ where} \quad U(s)=e^{\i s(H_{\text{mod}, A}+H_{\text{mod},\bar{A}})} \text{ and } \mathcal{O}\in \mathcal{A}_{A}.
\end{equation}
Note that only the two-sided operator $H_{\text{mod}}=H_{\text{mod}, A}+H_{\text{mod},\bar{A}}$ has a well-defined action on a state. In most cases the modular Hamiltonian is a complicated non-local operator. We consider examples where the modular Hamiltonian may be derived from the Rindler Hamiltonian and is thus known. The modular Hamiltonian then generates a generalised time evolution with the unitary $U(s)=e^{\i s(H_{\text{mod}, A}+H_{\text{mod},\bar{A}})}$. Similarly to the ordinary time evolution generated by the physical Hamiltonian, the modular time evolution is a symmetry of the subregion. Observers in $A$ and $\bar{A}$ then have the freedom to choose their modular time parameter.

We now discuss missing information for the parallel transport of intervals in CFTs on either boundary in the eternal AdS black hole geometry. The parallel transport operator $V_{\delta \lambda}$ for an interval $\lambda$ may be found by solving the modular-parallel transport equations \cite{Czech:2019vih},
\begin{equation}
    \begin{aligned}
        \partial_\lambda H_{\mathrm{mod}}-P_0^\lambda\left[\partial_\lambda H_{\mathrm{mod}}\right] & =\left[V_{\delta \lambda}(\lambda), H_{\mathrm{mod}}\right] \\
        P_0^\lambda\left[V_{\delta \lambda}(\lambda)\right] & =0,\label{eq:modular_Berry_transport}
    \end{aligned}
\end{equation}
where $P_0^\lambda$ is the projector into the fibre.
The Berry curvature then follows from
\begin{equation}
    R=[V_{\delta\lambda^i},V_{\delta\lambda^j}]d\lambda^i\wedge d\lambda^j.
    \label{eq:modular_Berry_Cruvature}
\end{equation}

The modular Berry curvature signals missing information in the presence of a black hole as follows. In particular, we consider a BTZ black hole with non-compact spatial direction and mass $M=\frac{c}{12}\left(\frac{2 \pi}{\beta}\right)^2$. There are two CFTs dual to the black hole. In each of these CFTs, we consider an interval with endpoints
$P^L_1:\left(-\frac{x}{2}, t_L=-t\right), \quad P^L_2:\left(\frac{x}{2}, t_L=-t\right), \quad P^R_3:\left(-\frac{x}{2}, t_R=t\right), \quad P^R_4:\left(\frac{x}{2}, t_R=t\right)$.
For the BTZ black hole with non-compact spatial direction and mass $M=\frac{c}{12}\left(\frac{2 \pi}{\beta}\right)^2$, the modular Hamiltonians for two disjoint intervals are given by \cite{Nakagawa:2018kvo}
\begin{equation}
    \begin{aligned}
        H_{12,\mathrm{mod}}&=\frac{\beta}{2 \pi \sinh \frac{\pi x}{\beta}} \int_{-\frac{x}{2}}^{\frac{x}{2}} d y\left(\cosh \frac{\pi x}{\beta}-\cosh \frac{2 \pi y}{\beta}\right) T_{z z}\left(-t+i \frac{\beta}{2}, y\right) \quad\mathrm{if}\, x>\frac{t}{2}\\
        H_{13,\mathrm{mod}}&=\frac{\beta}{2 \pi \cosh \frac{2 \pi t}{\beta}} \int_{-t+\frac{i \beta}{2}}^t d y\left(\sinh \frac{2 \pi t}{\beta}-\sinh \frac{2 \pi y}{\beta}\right) T_{z z}\left(y,-\frac{x}{2}\right) \quad\mathrm{if}\, x<\frac{t}{2}
    \end{aligned}
\end{equation}
and similarly for $ H_{34,\mathrm{mod}}$, $ H_{24,\mathrm{mod}}$. The Berry curvature may then be obtained by solving \eqref{eq:modular_Berry_transport} and \eqref{eq:modular_Berry_Cruvature}. The calculation was performed in \cite{Banerjee:2022jnv}. In particular, it was found that the Berry phase obtained from $ H_{13,\mathrm{mod}}$ vanishes if the time coordinates in the left and right boundary, $t_L$ and $t_R$, are aligned. On the other hand, a misalignment by $\delta=t_L-t_R$ yields the Berry curvature \cite{Banerjee:2022jnv}
\begin{equation}
    \hat{R}_{u_L, u_R}=-\frac{4 \pi^2}{\beta^2} \operatorname{sech}^2\left(\frac{\pi}{\beta}(2 t+\delta)\right) K_{+,12} d \delta \wedge d t,
\end{equation}
where $K_{+,12}$ is the generator of boosts for the modular Hamiltonian $ H_{\mathrm{mod}}=\int d\Sigma^{\mu}K^{\nu}T_{\mu\nu}$.

The Berry phase again signals a misalignment of the time coordinate between observers in an interval on the left and right boundary. The Berry phases vanishes only if their time coordinates are aligned. This misalignment is possible because an observer in the left boundary cannot communicate with an observer in the right boundary to align their time coordinates. Therefore, the choice of time coordinate in the boundary hidden behind the horizon presents missing information. The missing information is available in the global state which exhibits a relative phase due to the misaligned time coordinates similar to \eqref{eq:TimeEvolvedTwoSpinState}. However, the relative phase is not present in the reduced density matrix an observer in the boundary employs to obtain expectation values of observable.

\section{Conclusion and Outlook}

In the first part of this work we have established a relation between the presence of geometric phases in states of the Hilbert space and the possibility of consistently defining a trace on the corresponding operator algebra. For illustrative purposes, we have demonstrated this first for a simple two-spin model. Next we generalised the setup to two copies of infinitely many spins, prepared in a state that is the cyclic and separating vector for algebras of type III. We found that the trace is consistently defined if all of the geometric phases vanish. In this case, the aforementioned state is the cyclic and separating vector for algebras of type II. In this way, we provided a geometric argument, related to the geometry of entanglement, for the non-existence of the trace on type III algebras. Finally, we have discussed an explicit realisation of our result within holography, for the eternal black hole.

In the second part of this paper, we discussed how Berry phases signal missing information in system with and without entanglement. For systems without entanglement, considering the examples of a single spin in a magnetic field and of conformal transformations in a two-dimensional CFT, we discussed that the associated Berry phase signals the existence of global charges. This charge leads to additional phases in the global state inaccessible to a local observer. We then discussed missing information for Berry phases in entangled systems for the example of two intervals on a constant time slice in a two-dimensional CFT, and also for the two CFTs dual to the eternal black hole. In both cases, the Berry phase signals missing information related to independent symmetries in each subregion. These symmetries yield a relative phase in the global state describing the full system that cannot be measured by an observer restricted to a subregion.

These results exemplify the use of geometric phases for characterising the Hilbert spaces of both simple quantum systems and of quantum gravity in the AdS/CFT context. They quantify missing information that is also inherent in entangled systems and thus of central importance for the analysis of quantum Hilbert spaces based on von Neumann algebras.

Our results lead to several interesting follow-up research questions in relation to the interplay of geometric phases and entanglement. In the following, we elaborate on two of them: on implications for the Hawking-Page transition, as well as on the symmetry resolution of entanglement entropy.

\paragraph{Entanglement Geometry for the Hawking-Page Phase Transition}~\\
In quantum mechanics, it is straightforward to define
operators implementing a flow between different entanglement orbits. While local unitary operations of the form $U_L\otimes U_R$ do not alter the entanglement properties of a given state, linear combinations of local unitaries $\sum_ic_iU_L^{(i)}\otimes U_R^{(i)}$ generically change its entanglement properties by an amount depending on the coefficients $c_i$. Essentially, such operators may be understood as evolving the state with an interaction Hamiltonian. As an example, starting from a two-spin product state $|\!\downarrow\downarrow\rangle$ with vanishing entanglement, evolving this state using $\exp(\i H_{\text{int}}t)$, where $H_{\text{int}}=\gamma\,a_L^\dagger\otimes a_R^\dagger$ with the raising operators $a_{L/R}^\dagger=\frac{1}{2}(\sigma_{x,L/R}+\sigma_{y,L/R})$, results in an entangled state. The amount of entanglement depends on the duration $t$ of the evolution and the interaction strength $\gamma$ appearing in $H_{\text{int}}$. In particular, evolving for a duration of $t^\ast=\frac{1}{\gamma}$ results in a maximally entangled state. Since such operators change between entanglement orbits, they also change the geometric phase of the states that they act on. 

It will be interesting to study generalisations of this mechanism to infinite-dimensional systems in order to make contact to systems with operator algebras of type II and type III. In particular, we may think of a similar scenario for a black hole in holography. 
In view of describing the Hawking--Page transition \cite{Hawking:1982dh}
between empty AdS and the black hole geometry, we may 
simply start from  the CFT vacuum $|\text{vac}\rangle=|0_L\rangle|0_R\rangle$ and act with an operator of the form $\mathcal{O}\sim\exp\!\big(e^{-\beta\frac{E}{2}}a_L^\dagger a_R^\dagger)$  (cf. \cite{Israel:1976ur}). This also generates the entanglement between the left and right CFTs described by the TFD state. Since this operator changes the entanglement properties of the state, it also changes its geometric phase. It will be interesting to study how this operator is interpreted in the holographic setting.  The algebraic interpretation of the Hawking--Page transition is a change from two copies of type I$_\infty$ to two copies of type III$_1$ \cite{Leutheusser:2021frk,Leutheusser:2021qhd}. Analysing the geometric phase and its alteration for the Hawking--Page transition as described above will be useful for providing a geometric explanation for the transition related to the microscopic states of the system. In particular, due to the relation between geometric phases and entanglement, this analysis may put a new perspective on how the interior region of the eternal black hole emerges by increasing the entanglement.

From the algebraic perspective, the Hawking--Page transition is closely related to the Hagedorn transition. The Hagedorn transition was originally discovered in particle physics in the context of the confined and deconfined phases of quark matter \cite{Hagedorn:1965st}. In terms of operator algebras, the Hagedorn transition is understood as a transition from type I (confined phase) to type III (deconfined phase).\footnote{This phase transition was studied further in the context of string theory \cite{Atick:1988si} and also in ${\cal N}=4$ supersymmetric Yang--Mills theory \cite{Witten:1998zw,Sundborg:1999ue} in the context of holography.} Due to the similarity in terms of the operator algebraic description, understanding the Hawking-Page transition in terms of geometric phases may also be useful for analysing its microscopic origin, in analogy to the Hagedorn transition.

\paragraph{Entanglement Geometry and Symmetry Resolution}~\\
We have studied in detail how a geometric description of entanglement in bipartite quantum systems may be obtained using the SZK construction \cite{Sinolecka2002manifolds}. For each value of entanglement, analysing the Schmidt coefficients of the state given enables to define orbits of equal entanglement by quotienting the local unitary transformations with the appropriate stabiliser group. For each orbit, a connection may be defined, which by integration defines a geometric phase. In the light of our findings, it will be interesting to apply the SZK construction \cite{Sinolecka2002manifolds} to systems studied in the context of symmetry-resolved entanglement \cite{Goldstein:2017bua,Xavier:2018kqb}. Due to the symmetry resolution, the same orbit construction should be applicable to each charge sector. The geometric phase defined by the orbit construction may then be useful to interpret symmetry resolved entanglement in a geometric way. In particular, this method may be useful to gain a new interpretation for the equipartition of entanglement between sectors of different charge \cite{Xavier:2018kqb}. The equipartition of entanglement refers to the fact that systems in the thermodynamic limit, to leading order in the UV cutoff the symmetry-resolved entanglement entropy is independent of the charge. A recent study showed that, for a two-dimensional CFT with a global $\U(1)$ symmetry, when resolving w.r.t. the $\U(1)$ symmetry the equipartition of entanglement is related to the form of the $\U(1)$ characters \cite{DiGiulio:2022jjd}. In a different study it was shown that symmetry resolving a two-dimensional CFT w.r.t.~irreducible representations of the Virasoro group, equipartition of entanglement is related to the form of the conformal characters, or alternatively, the quantum dimension of the irreducible representation \cite{Northe:2023khz}. Finally in \cite{DiGiulio:2023nvz}, after establishing the symmetry resolution of elements of Tomita-Takesaki theory, it was found that also the modular correlation function of the charge density for a massless Dirac theory in two dimensions shows an equipartition when symmetry resolving w.r.t.~a global $\U(1)$ symmetry. In general however, at the time of writing, the origin of equipartition is not fully understood and requires further study. We expect that an analysis of equipartition in terms of geometric phases will put a new perspective on this question and in this way contribute to further demystify the equipartition of entanglement.

\acknowledgments

We thank Vijay Balasubramanian, Pablo Basteiro, Arpan Bhattacharyya, Saurya Das, Giuseppe Di Giulio, Ro Jefferson, René Meyer, Djordje Minic, Flavio Nogueira, Joris Raeymaekers, Shubho Roy, Eric Sharpe and Gideon Vos for useful discussions.

We acknowledge support by the Deutsche Forschungsgemeinschaft (DFG, German Research Foundation) under Germany's Excellence Strategy through the Würzburg-Dresden Cluster of Excellence on Complexity and Topology in Quantum Matter - ct.qmat (EXC 2147, project-id 390858490), via the SFB 1170 ToCoTronics (project-id 258499086) and via the  German-Israeli Project Cooperation (DIP) grant `Holography and the Swampland'. This research was also supported in part by Perimeter Institute for Theoretical Physics. Research at Perimeter Institute is supported by the Government of Canada through the Department of Innovation, Science and Economic Development and by the Province of Ontario through the Ministry of Research, Innovation and Science.

\appendix

\section{Some Essential Ingredients of von Neumann Algebras}
\label{app:EssentialAspectsvNAlgebras}

Here we briefly review the basic notions of vN algebras relevant for the discussion in the present paper. For more detailed reviews, we refer the reader to \cite{Witten:2018zxz,Witten:2021jzq,Sorce:2023fdx}.

The set of states describing a quantum system is called the Hilbert space. With these states, measurements of observables are performed by calculating expectation values of Hermitian operators representing the observables. Each such operator acts on the Hilbert space. The algebra satisfied by these operators is a vN algebra, i.e. a unital weakly closed $^\ast$-algebra of bounded operators. Equivalently, due to the bicommutant theorem \cite{vonNeumann1930algebra}, a vN algebra is a subalgebra of all bounded operators that is closed under the $^\ast$-operation and is its own double commutant. There exist three types of operator algebras, which we will briefly describe in the following.
\paragraph{Type I:}Algebras of type I are those typically encountered in quantum mechanics. They always allow for an irreducible representation. Equivalently, they allow for defining pure states on the algebra.\footnote{See e.g. \cite{Witten:2021jzq} for an explanation why this is equivalent. Note also that a state on the algebra is not the same as a state(vector) in the Hilbert space, the latter notion of a state being more common in quantum mechanics. Since in this paper, we will not explicitly need the concept of a state on the algebra, we will refer to the elements of the Hilbert space as states.\label{fn:PureStates}} A trace on the algebra can be naturally defined consistently, as we will discuss in an explicit example shortly in sec.~\ref{sec:IllustrativeExampleTwoSpin}. Utilising the trace, also density matrices and entanglement entropy are naturally defined. Type I algebras fall into two subclasses. For finite dimensional, i.e. $d$-level quantum systems such as qudits, the algebra is of type I$_d$, $d<\infty$. Those can always be understood as matrix algebras. For infinite dimensional quantum systems such as an infinite collection of qubits, the algebra is of type I$_\infty$. In this case, the trace on the algebra is not defined for every operator since e.g.~the trace of the identity diverges.
\paragraph{Type II:} Algebras of type II only appear for infinite-dimensional systems. They do not have an irreducible representation. This can be understood by the fact that an algebra of type II always has a commutant of the same type. The absence of an irreducible representation is equivalent to such algebras not allowing for pure states on the algebra (c.f. fn.~\ref{fn:PureStates}). However, a trace on the algebra, and therefore also density matrices, can be consistently defined. Again, there are two subclasses: type II$_1$ allows to define a trace for every operator. The other case is denoted as type II$_\infty$ and can be thought of as a tensor product of II$_1$ and I$_\infty$. For the same reason as with type I$_\infty$, the trace is not defined for every element of II$_\infty$.

Within the context of gravity, algebras of type II$_1$ and II$_\infty$ where found to appear for empty de Sitter spacetime \cite{Chandrasekaran:2022cip} and black holes in spacetimes of different (constant) curvature \cite{Witten:2021unn,Chandrasekaran:2022cip,Chandrasekaran:2022eqq}, respectively.

States corresponding to a type I algebra may contain a finite amount of entanglement, even for infinite dimensional quantum systems with a type I$_\infty$ description. This is not true for the case of type II algebras: in this case, entanglement is universally divergent since each state contains an infinite amount of entanglement. The same is true for algebras of type III, whose properties are stated in the following.
\paragraph{Type III:}As for type II, algebras of type III only appear in infinite dimensional systems and do not allow for an irreducible representation. Moreover, they do not allow for a consistent definition of a trace on the algebra. Correspondingly, density matrices and in particular the entanglement entropy are not defined. For this type, there exist three subclasses which can be characterised by the spectrum of the modular operator $\Delta$. If the only accumulation points of the spectrum of $\Delta$ are $0$ and $1$, the algebra is said to be of type III$_0$. If instead, the accumulation points are $0$ and the integer powers of some $\lambda^\ast<1$, one has type III$_{\lambda^\ast}$. If there are (at least) two such values $\lambda_1^\ast,\lambda_2^\ast\neq0$, the accumulation points are given by $(\lambda_1^\ast)^n(\lambda_2^\ast)^m,~n,m\in\mathds{Z}$. Therefore, the accumulation points can approximate any real number\footnote{This is true except for the case where $(\lambda_1^\ast)^n=\lambda^\p,(\lambda_2^\ast)^m=\lambda^\p,n,m\in\mathds{Z}$. If such a $\lambda^\p$ exists, the algebra is of type III$_{\lambda^\p}$. See e.g. \cite{Witten:2018zxz} for a more detailed discussion.} and the algebra is said to be of type III$_1$. It is the third case which is the typical scenario for quantum field theory and is the primary focus for the discussion in sec.~\ref{sec:TraceTypeIITypeIII}. Type III$_1$ also appears for the eternal black hole in AdS spacetime in the large $N$ limit \cite{Jefferson:2018ksk,Leutheusser:2021frk,Leutheusser:2021qhd}.
\paragraph{Factor:}An algebra ${\cal A}$ is referred to as a factor if its centre consists only of complex scalars, i.e.~if the operators $\mathcal{O}_C=z\1$, $z\in\mathds{C}$ are the only operators satisfying $[a,\mathcal{O}_C]=0$ for every $a\in{\cal A}$. In this case, the centre is said to be trivial.

\section{Entanglement Orbit Structure of Pure States}
\label{app:OrbitStructurePureStates}

Here we provide details on the SZK construction \cite{Sinolecka2002manifolds}.

Consider a generic quantum system in a Hilbert space $\cal H$ of size $d^2$. The pure states form the projective Hilbert space ${\cal H}_P=P(\mathds{C}^{d^2})=\CP{d^2-1}\subset{\cal H}$. Every state $|\psi\rangle$ can be written in the Schmidt decomposition
\begin{align}
    |\psi\rangle=\sum_{i=1}^d\kappa_i|i,\tilde{i}\rangle,\label{eq:schmidt_decomposed_state}
\end{align}
where the two bases $|i\rangle$, $|\tilde{i}\rangle$ imply some partition of the Hilbert space into sub-Hilbert spaces ${\cal H}\otimes\tilde{\cal H}$. The numbers $0\leq\kappa_i\leq1$ are known as Schmidt coefficients and, as the square roots of the diagonal entries of the reduced density matrix $\rho_{\text{red}}$ of either subsystem, uniquely fix the entanglement between the subsystems by
\begin{align}
    S_{\text{EE}}(|\psi\rangle)=-\tr(\rho_{\text{red}}\ln\rho_{\text{red}})=-\sum_{i=1}^d\kappa_i^2\ln\kappa_i^2.\label{eq:entanglement_entropy}
\end{align}
The three most interesting cases are
\begin{itemize}
    \item $\kappa_i=0~\forall i$ except for $\kappa_{i^\ast}=1$: vanishing entanglement,
    \item $\kappa_i=\frac{1}{\sqrt{d}}~\forall i$: maximal entanglement,
    \item $0\leq\kappa_1\leq\kappa_2\leq...\leq\kappa_d<1$: intermediate entanglement.
\end{itemize}
Note that the ordering of $\kappa_i$ in the third case is arbitrary. By a convenient labeling of the base vectors, the reduced density matrix can be written in the form
\begin{align}
    \rho_{\text{red}}=\text{diag}(\underbrace{0,...,0}_{m_0},\underbrace{\kappa_1,...,\kappa_1}_{m_1},...,\underbrace{\kappa_d,...,\kappa_d}_{m_d}),
\end{align}
such that the coefficients increase from the upper left to the lower right. Generically, several Schmidt coefficients may have the same value; above, this is denoted by the degeneracies $m_l$ for each Schmidt coefficient, i.e. the lowest non-vanishing Schmidt coefficient $\kappa_1$ appears $m_1$ times, the second lowest $\kappa_2$ appears $m_2$ times and so on. Note that this also includes $m_0$, which always indicates how many Schmidt coefficients are vanishing. By definition, $\sum_{l=0}^dm_l=d$. After finishing the general discussion of this construction, below we will give a minimal example using a two spin system for illustration.

The reduced density matrix $\rho_{\text{red}}$ is also understood as the square of the coefficient matrix of the Schmidt decomposition of $|\psi\rangle$,
\begin{align}
    |\psi\rangle=\sum_{i,j=1}^d\sqrt{\rho_{\text{red}}}_{ij}|i,j\rangle.
\end{align}
We can now ask for unitary matrices $U,~V\in\U(d)$ that, up to an overall phase, leave the Schmidt decomposition invariant,
\begin{align}
    U\otimes V^T|\psi\rangle=\sum_{i,j=1}^d(U\sqrt{\rho_{\text{red}}} V)_{ij}|i,j\rangle\sim|\psi\rangle.
\end{align}
It has been shown \cite{Sinolecka2002manifolds} that such matrices have the form
\begin{align}
    U=\begin{bmatrix}
        U_0 & 0 & 0 & \dots & 0 \\
        0 & U_1 & 0 &   &   \\
        0 & 0 & U_2 &   & \vdots \\
        \vdots &   &   & \ddots &   \\
        0 &   & \dots &   & U_d
    \end{bmatrix},\quad V=\begin{bmatrix}
        V_0 & 0 & 0 & \dots & 0 \\
        0 & U_1^\dagger & 0 &   &   \\
        0 & 0 & U_2^\dagger &   & \vdots \\
        \vdots &   &   & \ddots &   \\
        0 &   & \dots &   & U_d^\dagger
    \end{bmatrix}.
\end{align}
Here, $\dim(U_i)=\dim(V_i)=m_i$. This determines the orbit structure
\begin{align}
    \mathcal{O}_\psi=\frac{\U(d)}{\U(m_0)\times\U(m_1)\times...\times\U(m_d)}\times\frac{\U(d)}{\U(m_0)\times\U(1)},\label{eq:general_orbit}
\end{align}
which we give reason for in the following. The two factors of $\U(d)$ follow as the transformations consisting of local unitaries acting on the $|\psi\rangle$. Due to the matrices $U$ and $V$ leaving the Schmidt decomposition invariant, the first factor has the stabilisers $\U(m_i)$. Since $V_0$ is not determined by any $U_i$, the second factor also has the stabiliser $\U(m_0)$. Moreover, since $\rho_{\text{red}}$ has to be invariant only up to an overall phase, there is a factor $\U(1)$. This construction can be understood as a fibre bundle. The base space is formed by the reduced density matrices of the same spectrum. The fibre is formed by all pure states related by the partial trace to density matrices with a particular spectrum.

As a minimal example, consider two spins in the state
\begin{align}
    |\chi\rangle=\sin\sigma|\!\uparrow\downarrow\rangle+\cos\sigma|\!\downarrow\uparrow\rangle.
\end{align}
The projective Hilbert space for $|\chi\rangle$ is $\CP{3}$, so $d=2$ and there are only two Schmidt coefficients. Comparing with \eqref{eq:schmidt_decomposed_state} shows that $\kappa_1=\sin\sigma$ and $\kappa_2=\cos\sigma$. For particular values of $\sigma$, this leads to the following geometric classification:
\begin{align}
    \sigma=0: & \quad & \kappa_1=0,~\kappa_2=1 & \quad & \to S_{\text{EE}}=0\hspace{6.3cm}\notag\\[2pt]
    ~ & \quad & m_0=1,~m_1=1 & \quad & \mathcal{O}_\chi=\frac{\U(2)}{\U(1)\times\U(1)}\times\frac{\U(2)}{\U(1)\times\U(1)}=\CP{1}\times\CP{1}\notag\\[2pt]
    \sigma=\frac{\pi}{4}: & \quad & \kappa_1=\kappa_2=\frac{1}{\sqrt{2}} & \quad & \to S_{\text{EE}}=\ln2\hspace{5.9cm}\notag\\[2pt]
    ~ & \quad & m_0=0,~m_1=2 & \quad & \mathcal{O}_\chi=\frac{\U(2)}{\U(2)}\times\frac{\U(2)}{\U(1)}=\mathds{1}\times\RP{3}\hspace{2.9cm}\notag\\[2pt]
    0<\sigma<\frac{\pi}{4}: & \quad & 0<\kappa_1\neq\kappa_2<1 & \quad & \to0<S_{\text{EE}}<\ln2\hspace{5.15cm}\notag\\[2pt]
    ~ & \quad & m_0=0,~m_1=m_2=1 & \quad & \mathcal{O}\chi=\frac{\U(2)}{\U(1)\times\U(1)}\times\frac{\U(2)}{\U(1)}=\CP{1}\times\RP{3}\hspace{1.1cm}\notag
\end{align}
Note that for $\frac{\pi}{4}<\sigma<\frac{\pi}{2}$, the same situation as for $0<\sigma<\frac{\pi}{4}$ arises, just with the roles of $\kappa_1$ and $\kappa_2$ interchanged. Furthermore, $\sigma=\frac{\pi}{2}$ is the same as $\sigma=0$, except that $\kappa_1=1$ and $\kappa_2=0$. This reflects that there are two ways to embed $\CP{1}\times\CP{1}$ into $\CP{3}$, as mentioned below \eqref{eq:OrbitVanEnt} for bipartite quantum systems with general $d$.

\bibliographystyle{JHEP}
\bibliography{bib2.bib}

\end{document}